\documentclass[conference]{IEEEtran}
\ifCLASSINFOpdf
\else
\fi
\usepackage{times,amsmath,epsfig,subfigure,stmaryrd}
\usepackage{multirow}
\usepackage{alltt}
\usepackage{color}
\usepackage{xspace}
\usepackage{hhline}
\usepackage{soul}
\usepackage{comment}
\usepackage{wrapfig}
\usepackage[ruled,vlined,linesnumbered]{algorithm2e}
\usepackage{graphicx}
\usepackage[pdftex,colorlinks=true,citecolor=black,urlcolor=black,linkcolor=black]{hyperref}
\usepackage{xcolor}
\usepackage{listings}

\newcounter{enum}
\newenvironment{packed_enum}{
\begin{list}{\textbf{(\arabic{enum})}}{
  \setlength{\itemsep}{0pt}
  \setlength{\parskip}{0pt}
  \setlength{\labelwidth}{-4 pt}
  \setlength{\leftmargin}{0 pt}
  \setlength{\itemindent}{0pt}
  \usecounter{enum}}
}{\end{list}}

\newenvironment{packed_item}{
\begin{list}{\textbf{$\bullet$}}{
  \setlength{\itemsep}{0pt}
  \setlength{\parskip}{0pt}
  \setlength{\labelwidth}{-4 pt}
  \setlength{\leftmargin}{0 pt}
  \setlength{\itemindent}{4pt}
  \usecounter{item}}
}{\end{list}}

\definecolor{orange}{rgb}{1,0.5,0}
\definecolor{lightpurple}{rgb}{0.5,0,0.5}
\definecolor{purple}{rgb}{0.5,0,0.25}
\definecolor{mybrown}{rgb}{0.5,0,0}
\definecolor{darkgreen}{rgb}{0,0.5,0}

\newcommand{\hide}[1]{}
\newcommand{\system}{\textsc{IngestBase}\xspace}
\newcommand{\srm}[1]{\textcolor{red}{{\it Sam: #1}}}

\begin{document}
\bstctlcite{IEEEexample:BSTcontrol}

%
\title{\begin{huge}{\sc IngestBase}\end{huge}: A Declarative Data Ingestion System}

\author{\IEEEauthorblockN{Alekh Jindal}
\IEEEauthorblockA{Microsoft\\
aljindal@microsoft.com}
\and
\IEEEauthorblockN{Jorge-Arnulfo Quian\'e-Ruiz}
\IEEEauthorblockA{QCRI\\
jquianeruiz@qf.org.qa}
\and
\IEEEauthorblockN{Samuel Madden}
\IEEEauthorblockA{MIT\\
madden@csail.mit.edu}}


%


\maketitle

\begin{abstract}
Big data applications have fast arriving data that must be quickly ingested.
At the same time, they have specific needs to preprocess and transform the data before it could be put to use.
The current practice is to do these preparatory transformations once the data is already ingested, however this is expensive to run and cumbersome to manage.
As a result, there is a need to push data preprocessing down to the ingestion itself.
In this paper, we present a declarative data ingestion system, called \system{}, to allow application developers to plan and specify their data ingestion logic in a more systematic manner.
We introduce the notion of \textit{ingestions plans}, analogous to query plans, and present a declarative ingestion language to help developers easily build sophisticated ingestion plans.
\system{} provides an extensible ingestion optimizer to rewrite and optimize ingestion plans by applying rules such as operator reordering and pipelining.
Finally, the \system{} runtime engine runs the optimized ingestion plan in a distributed and fault-tolerant manner.
Later, at query processing time, \system{} supports ingestion-aware data access and interfaces with upstream query processors, such as Hadoop MapReduce and Spark, to post-process the ingested data.
We demonstrate through a number of experiments that \system: 
(i)~is flexible enough to express a variety of ingestion techniques, 
(ii)~incurs a low ingestion overhead, 
(iii)~provides efficient access to the ingested data, and 
(iv)~has much better performance, up to $6$ times, than preparing data as an afterthought, via a query processor.
\end{abstract}


%
\IEEEpeerreviewmaketitle

\section{Introduction}
\label{sec:introduction}

Modern big data applications witness massive amounts of continuously and quickly arriving data.
At the same time, this data needs to be preprocessed and optimized in a specific manner before it could be put to use.
Examples of such applications include: 
(i)~data exploration over periodically arriving scientific datasets, e.g.,~astronomy;
(ii)~analyzing service logs, e.g.,~from cloud services, which need to be quickly ingested for real time debugging;
(iii)~approximate query processing over fast arriving social network data, e.g.,~tweets, to get the latest trends;
(iv)~quality checking and cleaning commodity data, e.g.,~news content, before selling it on the data market; and
(v)~archiving high velocity telecom data, e.g.,~phone calls, for security purposes.
In all these scenarios, data needs to be consumed as soon as it arrives and so once it gets ingested there is little room for further preprocessing, which is anyways prohibitively expensive due to the massive data volumes.

As a result, applications developers need to carefully design their data ingestion pipelines and 
push the application specific data preparation logic down to data ingestion itself.
For instance, applying multi-dimensional partitioning to slice and dice the data in different ways for data exploration, or detecting and fixing data quality violations for data market applications, or considering different erasure codes to reduce the storage footprint during data archiving.
This is in contrast to static and hard-coded data ingestion pipelines in traditional databases as well as in big data systems like Hadoop.
Hadoop Distributed File System (HDFS), for instance, chunks input data into fixed sized blocks, replicates each block (three times by default), and stores them on different machines for fault tolerance.
Some efforts have tried to add additional steps to this static pipeline, such as indexing~\cite{hail}, co-partitioning~\cite{cohadoop}, and erasure coding~\cite{hdfs-raid}.
However, each of these forks out a new storage system with one additional feature at a time and does not offer full flexibility to specify arbitrary application-specific ingestion logic.

The current practice to deal with application-specific ingestion needs is to additionally deploy so called \textit{cooking jobs} to prepare the data, i.e.,~use a query processor to run preprocessing jobs once the data is already ingested.
However, this forces the users to spend additional time and money, as well as introduces another dependency before the data could be put to use.
Cooking jobs are also hard to share because they contain custom ad-hoc logic not necessarily understood by others, i.e.,~they lack a formal data ingestion language.
In addition, we now have both the ingested as well as the cooked data at the same time, i.e.,~data duplication, and in a fault-tolerance manner (e.g.,~replicated).
Finally, the cooking jobs end up overloading the compute clusters, even when data ingestion typically runs on separate capacity which is often underutilized.
Thus, with cooking jobs, the users end up creating additional data pipelines, which are tedious to build, expensive to run, and cumbersome to manage.

In this paper, we identify data ingestion as an explicit step that needs to be specified and planned in a more systematic manner.
We present \system{}, a flexible and declarative data ingestion system to quickly prepare the incoming data for application specific requirements.
At the same time, \system{} hides the ingestion processing complexity from the users, similar to databases hiding the query processing complexity.
To do this, \system{} exposes a declarative language interface, the \textit{ingestion language}, to easily express arbitrary ingestion logic, essentially an operator DAG connecting raw data sources to application ready data in the storage system.
\system{} uses an optimizer to rewrite and compile declarative ingestion statements into an efficient ingestion plan.
\system{} has a runtime engine that runs the optimized ingestion plan in a distributed and fault-tolerant manner over a cluster of machines.
Finally, \system{} provides a data access kernel to support ingestion aware query processing via higher level substrates, such as MapReduce and Spark.

In summary, our key contributions are as follows:

\begin{packed_enum}
\item We introduce the notion of \textit{ingestion plans}, analogous to query plans in relational databases, to specify a sequence of transformations that should be applied to raw data as it is ingested into a storage system. To easily build ingestion plans, we describe a declarative data ingestion language to express complex ingestion logic by composing a variety of ingestion operators and their data flow (Section~\ref{section:uploadplan}). 
\item We present an extensible, rule-based ingestion optimizer to rewrite and optimize ingestion plans, via techniques such as operator reordering and pipelining.
We further describe the \system{} runtime engine to efficiently run the optimized ingestion plan in a distributed and fault-tolerant manner.
In particular, we show how the system allows users to control the fault-tolerance mechanism for their data based on their ingestion plans (Sections~\ref{section:optimizer} and~\ref{section:execute}).
\item We describe the \system{} support for ingestion-aware data access, i.e., leveraging the ingest processing for efficient data access via upstream query processors.
Specifically, we show how our prototype implementation works with two storage-compute combinations, namely HDFS-MapReduce and HDFS-Spark.
(Sections~\ref{section:dataaccess} and~\ref{section:prototype}).
\item Finally, we present experimental evaluation over TPC-H dataset to:
(i)~show the overhead of \system{} and compare it with plain HDFS upload times,
(ii)~compare the effectiveness of ingest-aware query processing compared to both MapReduce and Hive,
(iii)~contrast \system{} with ingestion via cooking jobs (using Hive), and
(iv)~show the fault-tolerance behavior of \system{} (Section~\ref{section:experiments}).
\end{packed_enum}

In the remainder of the paper, we first discuss four different case study scenarios to understand the data ingestion pain in modern applications (Section~\ref{section:ingestionpain}).
Then, we describe an overview of the \system architecture (Section~\ref{section:abstraction}) before presenting our core contributions (Sections~\ref{section:uploadplan}--~\ref{section:experiments}).

\section{The Data Ingestion Pain}
\label{section:ingestionpain}

Let us first see the data ingestion pain in modern applications
and understand the need for a declarative ingestion system.
Below we describe four case study scenarios, namely data cleaning, data sampling, data analytics, and data storage, to highlight the 
preprocessing needs in modern big data applications and motivate a more systematic approach to it.


\subsection{Data Cleaning}
\label{section:cleaning}

Data cleaning is traditionally done as an afterthought, i.e.,~the data cleaning process starts once a dataset has already been uploaded~\cite{nadeef,holistic,bigdansing}. 
This means that users have to apply tedious and time consuming data cleaning transformation before the data could be put to use.
In contrast, cleaning data while ingesting the datasets would speed up the entire cleaning process. Users want to detect the portions of the data that violate their business rules, often an expensive step in the data cleaning process~\cite{holistic}, and apply simple repairs. Though data repair may be an iterative process, users could apply one-pass repairs on their datasets.  Below, we discuss examples of a few data cleaning operations.

\noindent\textbf{Functional Dependency Checks.} Consider the TPC-H lineitem table, which includes one entry per item per order in a business analytics application.  This table includes a shipdate field (the date the item shipped) and a linestatus field (whether the item has shipped or not). We may want to enforce a functional dependency (FD) that \textit{shipdate determines linestatus}, i.e., products shipped on the same date have the same linestatus. 
This would require to partition the data on shipdate, iterate over every pair of tuples in each partition, and check whether or not there is a functional dependency violation. Subsequently, the violated records could be output to a violations file (for further correction) or even discarded.

\noindent\textbf{Denial Constraint Checks.} Consider again the TPC-H lineitem dataset.
Suppose the user wants to check the following denial constraint (DC): {\it each item sold in quantity less than $3$ does not have a discount of more than $9\%$}.
This requires to scan the lineitem table, check each tuple for this denial constraint, and then store both the violating tuples as well as the original data.

\noindent\textbf{Single-pass Repair.} Besides detecting violations to data quality rules, users may also want to perform single-pass repairs. For example, consider a tax dataset having a country\_code attribute. 
In case the country code is not valid, users may want to correct the code using a dictionary (e.g.,~changing a value ``mexico'' to its corresponding code ``MX''). 
This would require parsing the country\_code attribute in the dataset, checking if the code is valid, and looking up in the dictionary in case the code is invalid.
Only the corrected values are finally stored.

\subsection{Data Sampling}

Sampling is a common technique to gather quick insights from very large datasets~\cite{blinkdb}. Samples can be used to quickly evaluate statistical properties of data (e.g., approximate averages or counts in certain subgroups), or to get a representative subset of the data. A key problem in using samples is the process of generating samples themselves:~producing a sample requires an entire pass over the data. Rather, the users want to collect samples as the data is being ingested, with minimal overhead. We discuss a few scenarios below.

\noindent\textbf{Random Sampling.} Users may want to create Bernoulli samples by probabilistically replicating some of the tuples in the dataset 
and collecting them into a separate physical file, i.e., in addition to collecting all tuples into a base file anyways.
Likewise, users may also want to create reservoir samples by adding each tuple into a reservoir, removing tuples from it with a given probability, and then finally emitting the reservoir as samples in the end.

\noindent\textbf{Stratified Sampling.} Besides pure random sampling, users may also generate stratified samples, where rare subgroups are over-represented vs common sub-groups. For example, in a dataset about people by state, a larger fraction of records from North Dakota might be included than from California, to ensure that enough records about North Dakota are present to achieve a target level of statistical confidence. Such samples are commonly used in databases to produce statistical approximations~\cite{congressional-sampling,blinkdb}. 
This requires to partition the data on the stratification attribute and randomly pick records from each strata (partition). The number of records picked from each partition is proportional to the partition size.

\subsection{Data Analytics}
\label{subsection:dataanalytics}
Data analytics often requires special data formats for good performance.
The typical practice is to either create these formats once the data has been already uploaded to a storage system~\cite{hadooppp}, 
or to modify the storage system with the application-specific logic~\cite{hail,cohadoop}.
In contrast, the developers would want to simply specify their formats (declaratively) and let the ingestion system take care of creating appropriate files in the storage system.

\noindent\textbf{Co-partitioning.} Users may want to apply custom data partitioning when ingesting data. For example, users can mimic CoHadoop~\cite{cohadoop}, where two data sets with a common ``join'' attribute are stored together, enabling efficient joins without repartitioning. 
Users could further sample the two relations and evaluate the skew in the join attribute, creating more balanced co-partitions.

\noindent\textbf{Layouts \& Indexes.} Users may want to plug-in alternate data layouts and indexes, e.g., RCFile~\cite{rcfile} (the default Hive~\cite{hive} layout in HDFS), Trojan Layouts~\cite{trojanlayout} (i.e.,~a different data layout for each data replica) and Hail~\cite{hail} (i.e.,~a different index for each data replica).
Creating Trojan Layouts, for instance, would require to create data blocks, replicate each block three times, and serialize each block replica differently (e.g., row, column, and RCFile).
Users may also want to create different layouts for different parts of data replicas, i.e., sub-divide the data blocks within a data replica and create a different layout for each of them.
Such hybrid replicas improve query robustness as more queries are likely to see at least some of the data blocks in favorable layouts.


\noindent\textbf{Data Placement.}  
With large data centers, users may want to control how the data is placed in them, e.g., placing hot and cold data blocks differently.
This requires looking at the contents of each data block when placing it into a cluster.
Such content based data placement could be further useful for: (ii)~improving data locality, (iii)~isolating concurrent queries to different nodes, and (i)~utilizing a portion of the cluster to save energy or to multiplex resources.

\subsection{Data Storage}
Despite the plummeting price of disks, storage space still remains a concern in replicated storage systems with  large datasets.
Below we describe two scenarios on how users may want to optimize the storage space. 

\noindent\textbf{Replicated Storage.} 
Users may want to control both what parts of the data are replicated and how many times. This control becomes crucial when different parts of the data have different relative importance. For example, a user storing weblogs might replicate the most recent logs (hot data) more frequently for higher availability, compared to the massive older logs (cold data). 
This would require partitioning the data on date (could be trivial in case of time series), and then applying the replication selectively.

\noindent\textbf{Erasure-coded Storage.} Erasure coding is an alternative to replication for handling failures. The advantage of erasure coding over replication is that it provides the same degree of redundancy as replication at lower storage overhead (but with a higher access cost in the event of failure).
Creating an arbitrary erasure code would require dividing the input blocks into stripes and applying erasure coding for each stripe.
As with data layouts and indexes, user may want to use different erasure codes, or a mix of replication and erasure codes, for different portions of the data (e.g., erasure codes for cold data and replication for hot data).
The recovery mechanisms, however, should work with both erasure codes and replication.

\subsection{Remarks}
We see that the above applications require custom data ingestion logic, i.e.,~users would like to provide different cleaning rules, sampling techniques, data layouts, and erasure codes.
Employing cooking jobs for such applications is tedious, time taking, and inefficient.
Rather the users would want to specify how the data should be transformed as it gets ingested, without incurring additional cooking jobs or worrying about the low-level details.
Thus, we need a systematic and declarative data ingestion system.

%
%
%
%

\section{\system{} Overview}
\label{section:abstraction}

\begin{figure}[!t]	
\centering
\includegraphics[width=2.4in]{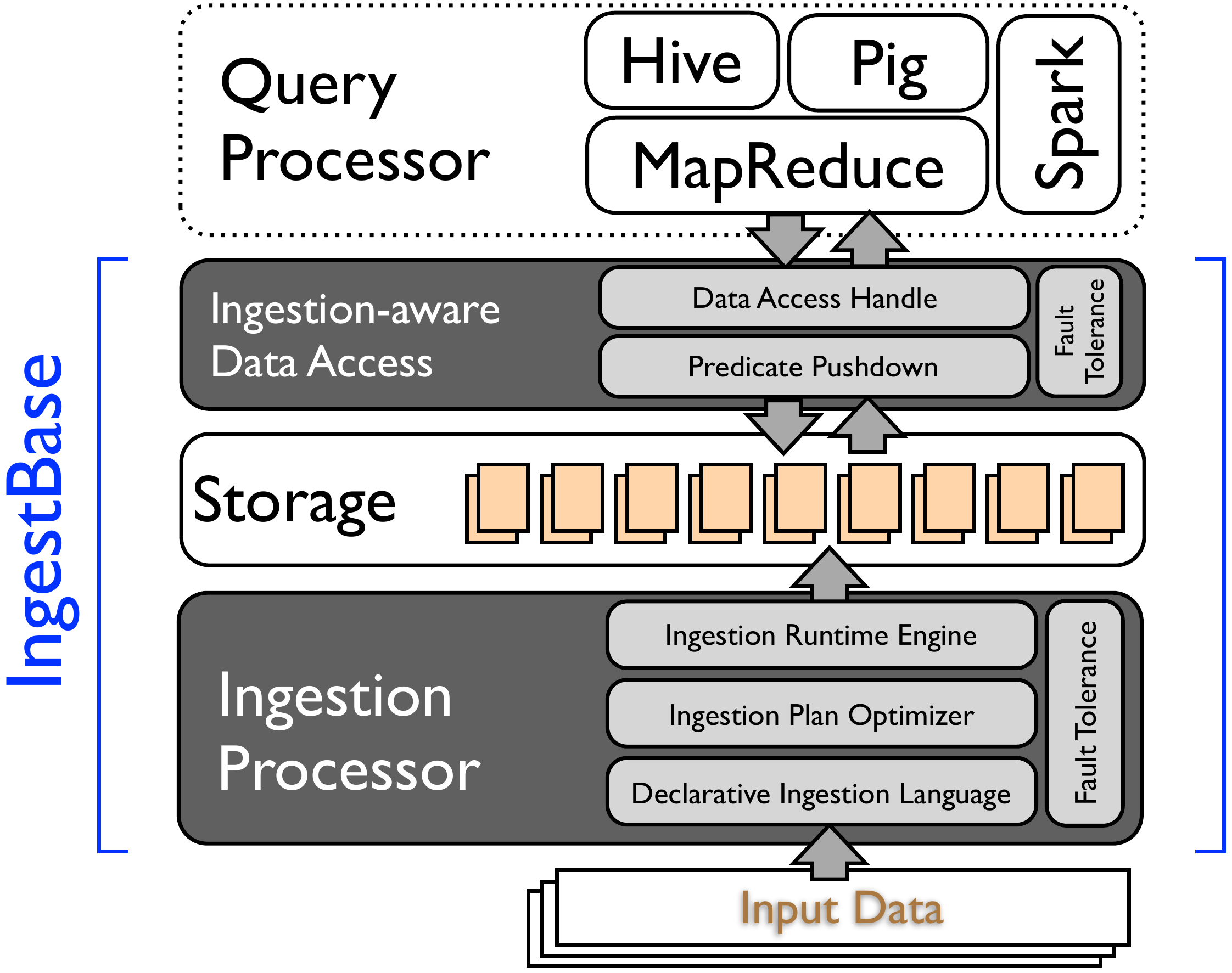}
\vspace{-0.3cm}
\caption{IngestBase System Architecture}
\label{figure:overview}
\vspace{-0.5cm}
\end{figure}

The goal of \system{} is to allow developers to easily express and efficiently run arbitrary data ingestion logic.
In our earlier works, we demonstrated flexible data upload to HDFS~\cite{cartilage}, as well as we showed how it could be used for scalable violation detection~\cite{bigdansing} and robust data partitioning~\cite{amoeba,ShanbhagJNMQAE}.
This paper describes a full-fledged, declarative data ingestion system that could work with arbitrary storage and query processor substrates.
Figure~\ref{figure:overview} illustrates the architecture of our system.

The input to be ingested using \system is represented as data items, referred to as \textit{ingest data items}.
At the very beginning, the ingest data items are simply the raw input files.
However, these could be later broken into smaller ingest data items, such as file chunks or records for fine-grained ingestion logic, e.g., applying chunk level replication or detecting null values in each record.
Each ingest data item is further associated with a list of \textit{labels} denoting its lineage during ingestion.
Finally, \textit{ingestion operators} specify the logic to transform the ingest data items, i.e.,~$\texttt{IngestOp}:\texttt{LID}\rightarrow \texttt{LID'}$, where \texttt{LID} and \texttt{LID'} are input and output set of labelled ingest data items. 
For example, the single-pass repair from Section~\ref{section:cleaning} would output only a repaired tuple, i.e.,~$\texttt{SinglePassRepair}:\texttt{t}\rightarrow \texttt{t}_{\text{repaired}} | \phi$.
Ingestion operator follows the iterator model with the following API:
\begin{packed_item}
\item \textit{initialize}: initialize an operator for the first time.
\item \textit{setInput}: assign the set of input ingest data items.
\item \textit{hasNext}: check whether next output is available.
\item \textit{next}: get the next output labelled ingest data item.
\item \textit{finalize}: cleanup the ingestion operator in the end.
\end{packed_item}

With ingest data items and the ingest operators as the building blocks, \system{} allows to create arbitrary operator DAGs, called \textit{ingestion plans}. 
An ingestion plan can further control the data flow by selectively choosing which ingest data items go to which portions of the DAG.
\system{} makes it easier for the users to build sophisticated ingestion plans by providing a declarative ingestion language. The ingestion plan is then optimized via the \textit{ingestion optimizer}, which choses to push-down or push-up the ingestion operators, pipeline the data flow across several ingest operators, and block the data flow wherever needed. Finally, the \system{} runtime engine runs the resulting optimized ingestion plan in a distributed and fault-tolerant manner.

\section{Data Ingestion Language}
\label{section:uploadplan}

In this section, we describe the declarative ingestion language in \system{}. In contrast to the current practice of using query processing language to cook the data, to the best of our knowledge, this is the first work to propose primitives for an ingestion language. There are two parts to our ingestion language: (i)~the declarative ingestion operators to specify the application-specific data transformation during ingestion, and (ii)~the declarative ingestion data flow to control (via the use of labels) which data items flow through different parts of the ingestion plan. We describe these two below.

\subsection{Ingestion Operators}

The ingestion operators help address three ingestion needs: \textit{what} to ingest, \textit{where} to ingest, and \textit{how} to ingest, similar to the what, where, and how of data storage proposed in~\cite{wwhow}.
For a given application, these ingestion needs could be derived using storage optimizer tools~\cite{wwhow, rodentstore}.
Users can then define \textit{what to ingest} using a \texttt{SELECT} statement as follows:

\vspace{-.2cm}
\begin{scriptsize}
\begin{verbatim}
s1 = SELECT projection
      FROM LID USING parser
      WHERE filter
      REPLICATE BY replicator;
\end{verbatim}
\end{scriptsize}
\vspace{-.2cm}

\noindent While the above syntax is very similar to standard SQL select statement (except replication and result assignment), the \textit{projection}, \textit{parser}, \textit{filter}, and \textit{replicator} can also be provided as custom ingest operators. For instance, we may project machine learning features for each tuple or replicate the ingest data items probabilistically. On compilation, the ingestion operators in the \texttt{SELECT} statement are chained as follows:\\
\indent $\textit{LID}\rightarrow\textit{parser}\rightarrow\textit{filter}\rightarrow\textit{projection}\rightarrow\textit{replicator}$\\
For a \texttt{SELECT} statement to be valid, the output and input ingest data items of consecutive ingest operators must match, i.e., they should have the same granularity and the same schema.
Next we show the \texttt{FORMAT} statement to describe \textit{how to ingest} the data.

\vspace{-.2cm}
\begin{scriptsize}
\begin{verbatim}
s2 = FORMAT s1
      PARTITION BY partition
      CHUNK BY chunk
      ORDER BY order 
      SERIALIZE AS serializer;
\end{verbatim}
\end{scriptsize}
\vspace{-.2cm}


\noindent 
Operators in the above \texttt{FORMAT} statement are chained in the order in which they appear, i.e.,~the chaining in $s2$ is $\textit{partition}\rightarrow\textit{chunk}\rightarrow\textit{order}\rightarrow\textit{serialize}$. Users can create alternate chains by changing the order of these operators, e.g., ordering before chunking will create a global sort order as opposed to per-chunk sort order in $s2$. The operators in \texttt{FORMAT} statement could also appear multiple times, e.g.,~users could apply multi-level partitioning (across and within chunks) as follows:

\vspace{-.2cm}
\begin{scriptsize}
\begin{verbatim}
s3 = FORMAT s1
      PARTITION BY top-level-partition
      CHUNK BY chunk
      PARTITION BY intra-chunk-partition
      ORDER BY order 
      SERIALIZE AS serializer;
\end{verbatim}
\end{scriptsize}
\vspace{-.2cm}

Finally, \textit{where to ingest} is specified using the following \texttt{STORE} statement:

\vspace{-.2cm}
\begin{scriptsize}
\begin{verbatim}
s4 = STORE s3
      LOCATE USING locator
      UPLOAD TO target;
\end{verbatim}
\end{scriptsize}
\vspace{-.2cm}

\noindent The \textit{locator} operator specifies which ingest data items must be co-located (or anti-located), while the \textit{target} operator specifies the final storage substrate. Note that \textit{target} only points to the registered storage location; the actual binding of \system{} with the storage system is a bit more involved, as described in Section~\ref{section:prototype}.

\begin{figure*}[!t]	
\hspace{-0.2cm}
\subfigure[Example Log Ingestion Plan.]{
\includegraphics[height=2.1in]{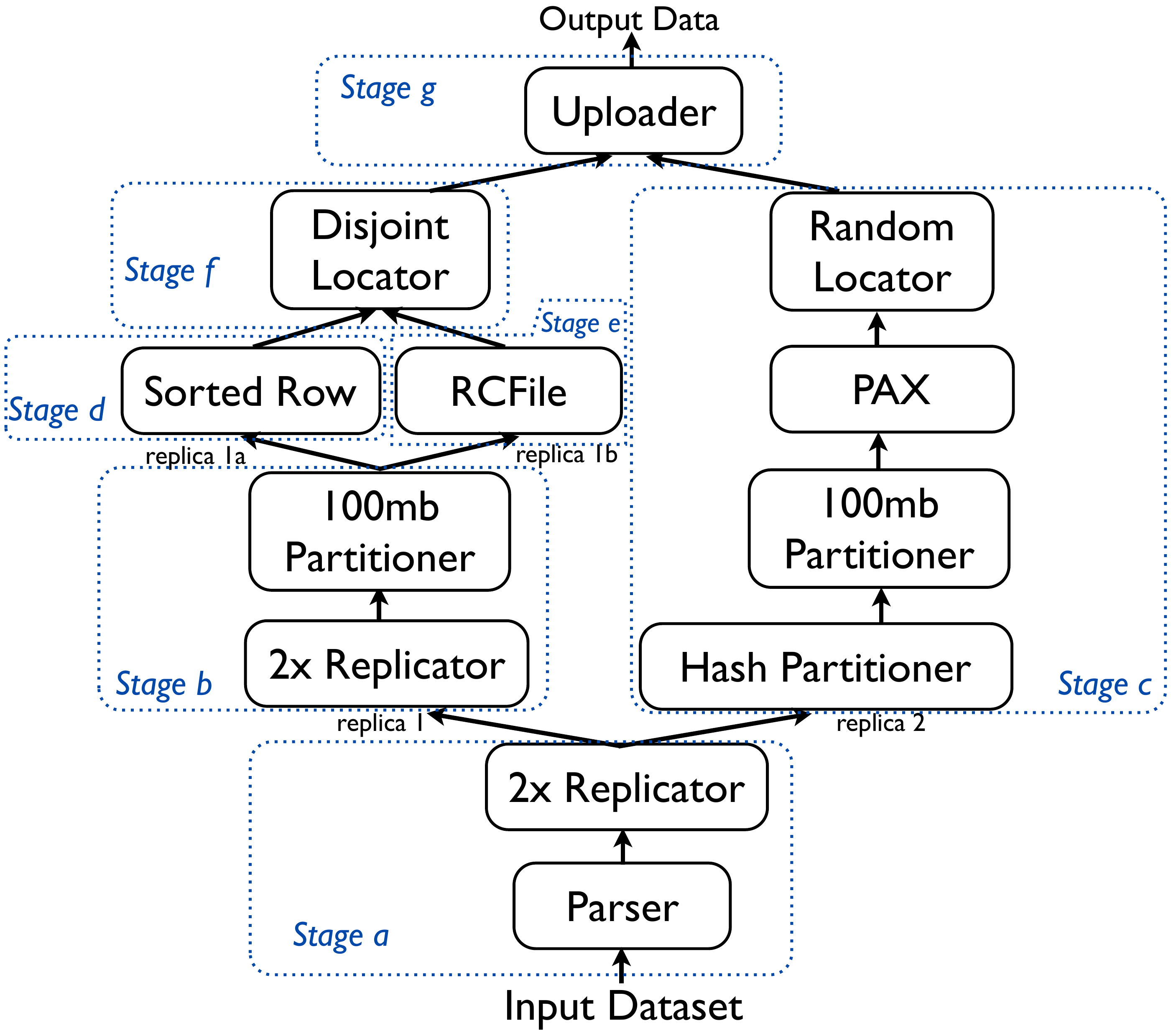}
\label{figure:logIngestionExample}
}
\hspace{0.1cm}
\subfigure[Operator reordering.]{
\includegraphics[height=2.1in]{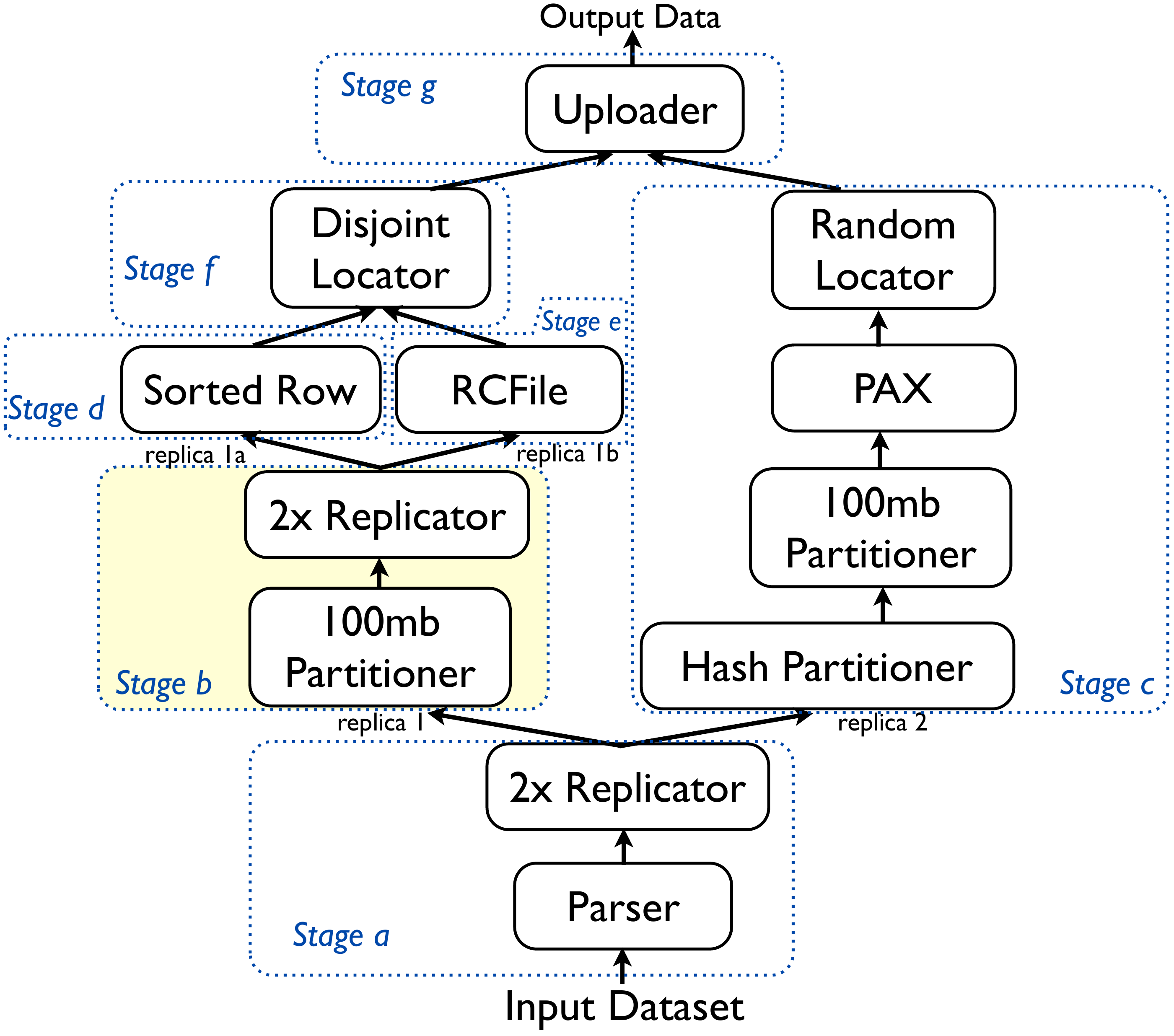}
\label{figure:logIngestionExampleReordering}
}
\hspace{0.1cm}
\subfigure[Operator pipelining.]{
\includegraphics[height=2.1in]{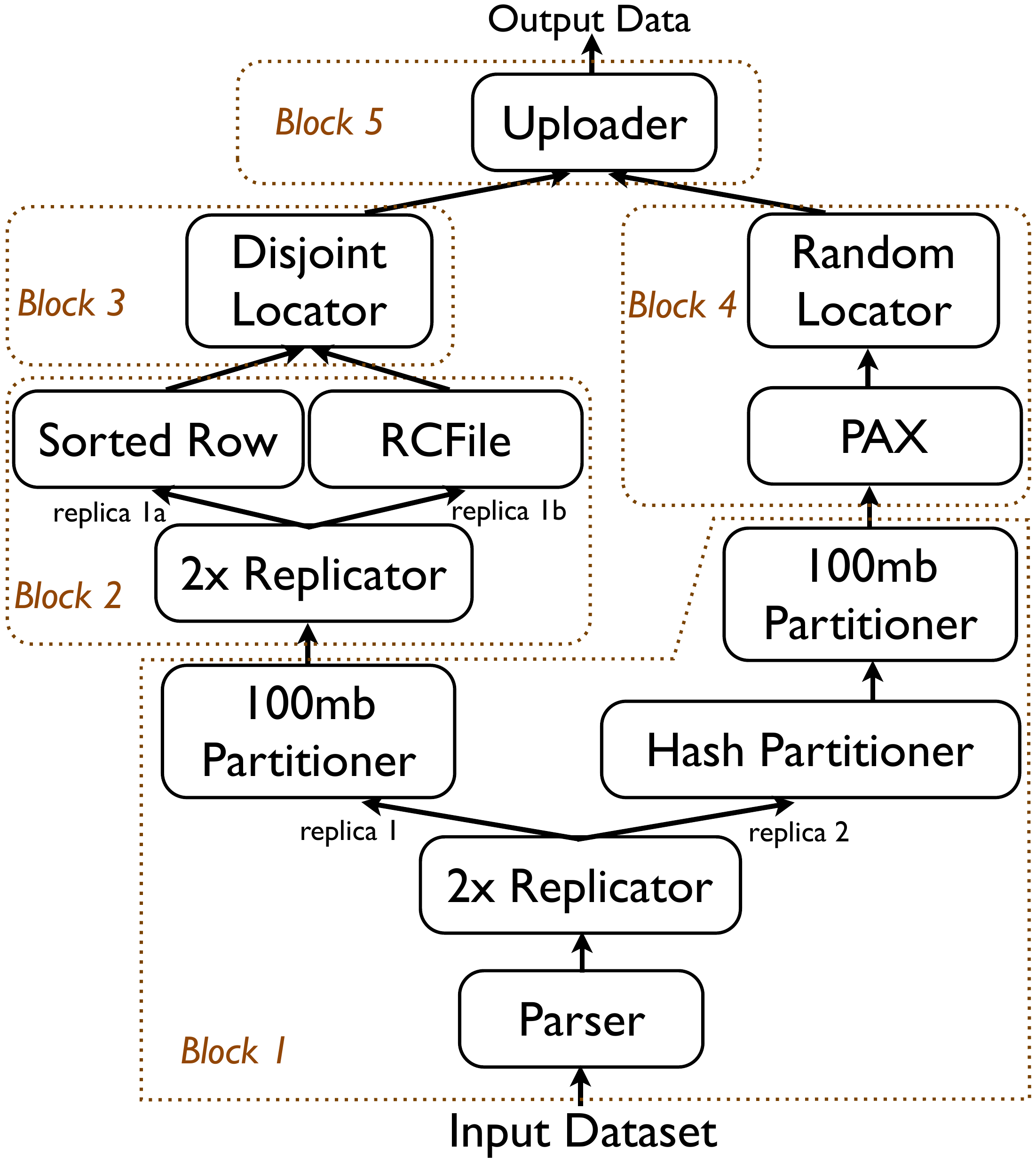}
\label{figure:logIngestionExamplePipelining}
}
\vspace{-0.3cm}
\caption{Illustrating ingestion plan and its optimization.}
\label{figure:IngestionPlanIllustration}
\vspace{-0.4cm}
\end{figure*}

\subsection{Ingestion Dataflow}

In the previous section, we saw the declarative statements for specifying and chaining ingestion operators.
Our ingestion language further allows the users to control the ingestion data flow, i.e., selectivity feed different portions of the ingest data items to different ingest operators.
To do so, we define a \textit{data flow stage} as a set of ingest operators operating on a set of ingest data items.
Recall, that the ingest data items have an associated set of labels denoting the transformations applied to them so far.
We use these labels to filter the relevant data items for each stage.

\vspace{-.2cm}
\begin{scriptsize}
\begin{verbatim}
CREATE STAGE a
  USING s1,s2,..,sm
  WHERE l_op1=v1,l_op2=v2,..,l_opn=vn
\end{verbatim}
\end{scriptsize}
\vspace{-.2cm}

\noindent In the above, we define a stage \texttt{a} with the ingest operators in $s_1-s_m$ (which could be either of the \texttt{SELECT}, \texttt{FORMAT}, and \texttt{STORE}) and operating on ingest data items that have labels $\text{l}_{\text{op}_i} = \text{v}_i$. 
For example, consider ingesting hourly data and assume that the \texttt{parser} operator assigns the file creation timestamp as the label for each ingest data item, the following stage ingests only the last hour of data each time:

\vspace{-.2cm}
\begin{scriptsize}
\begin{verbatim}
s1 = SELECT * FROM input;
CREATE STAGE a 
  USING s1 
  WHERE l_parser > now-1;
\end{verbatim}
\end{scriptsize}
\vspace{-.2cm}


\noindent Multiple stages could be chained to each other using the \texttt{CHAIN STAGE} statement as shown below:

\vspace{-.2cm}
\begin{scriptsize}
\begin{verbatim}
CHAIN STAGE b TO a1,a2,..,ak
  USING s1,s2,..,sm
  WHERE l_op1=v1,l_op2=v2,..,l_opn=vn
\end{verbatim}
\end{scriptsize}
\vspace{-.2cm}

\noindent Note that the above statement performs a union all on the outputs from stages $a_1,a_2,..,a_k$ before feeding it to stage $b$.

\vspace{0.1cm}
By defining stages on top of the ingestion operators, users can selectively process different ingest data items in different parts of the ingestion plan.
Such selective ingestion capability is useful for: 
(i)~handling heterogeneous data where different portions of the data have different characteristics and hence they need different data ingestion logic,
(ii)~supporting multiple workload types, e.g., graph and relational analytics, each requiring the data to be shoehorned differently, and
(iii)~reducing the risk of picking the wrong ingestion logic, e.g., due to changes in the workload, by applying multiple logic in the first place.

%
%


\subsection{Example: Log Analytics}
\label{sec:logIngestionExample}

Let us now illustrate our language via an example.
Consider a log analytics scenario where large volumes of logs are collected from a cloud service.
These logs need to ingested quickly with a low overhead.
Later, in case of any problems with the cloud service, e.g., disruption or slow performance, the service administrators need to quickly search the relevant log lines.
Each log line contains a combination of structured (e.g., timestamp, machine name) and unstructured (e.g., the error stack, manual user commands) data items.

For such an application, developers may create the following ingestion logic: create three data replicas and apply a different set of operators to each of them; first two of the three replicas differ only in their layout (sorted row and RCFile), while the third replica uses logical partitioning in addition to the physical partitioning. As a result, the first two replicas are suitable for selection and projection queries, while the third replica is suitable for join and aggregation queries. The ingestion statements for these are as follows:

\vspace{-.2cm}
\begin{scriptsize}
\begin{verbatim}
s1 = SELECT * FROM input USING parser REPLICATE BY 2;
s2 = SELECT * FROM s1 REPLICATE  BY 2;
s3 = FORMAT s2 CHUNK BY 100mbBlocks;
s4 = FORMAT s3 SERIALIZE AS sortedRow;
s5 = FORMAT s3 SERIALIZE AS rcFile;
s6 = FORMAT s1 PARTITION BY hash CHUNK BY 100mbBlocks
     SERIALIZE AS pax;
s7 = STORE s4,s5 LOCATE USING disjointLocator;
s8 = STORE s6 LOCATE USING randomLocator;
s9 = STORE s7,s8 UPLOAD TO hdfsStorage;
\end{verbatim}
\end{scriptsize}
\vspace{-.2cm}

\noindent The corresponding ingestion data flow is described as follows:

\vspace{-.2cm}
\begin{scriptsize}
\begin{verbatim}
CREATE STAGE a USING s1;
CHAIN STAGE b TO a USING s2,s3 WHERE l_replicate1=1;
CHAIN STAGE c TO a USING s6,s8 WHERE l_replicate1=2;
CHAIN STAGE d TO b USING s4 WHERE l_replicate2=1;
CHAIN STAGE e TO b USING s5 WHERE l_replicate2=2;
CHAIN STAGE f TO d,e USING s7;
CHAIN STAGE g TO c,f USING s9;
\end{verbatim}
\end{scriptsize}
\vspace{-.2cm}

Finally, Figure~\ref{figure:logIngestionExample} depicts the resulting log ingestion logic, as described above.
As also noted in~\cite{wwhow}, we see that thinking in terms of \textit{what}, \textit{where}, and \textit{how} in \system{} makes it more intuitive to reason about arbitrary data ingestion operations.
Also, the data flow primitives in \system{} allow users to easily control and selectively process the ingest data items.
Together, the declarative ingestion operators and the data flow primitives allow users to quickly stitch sophisticated ingestion plans for their applications.

\section{Ingestion Optimizer}
\label{section:optimizer}


\system{} takes the declarative ingestion statements and compiles them into a DAG, the ingestion plan, as shown in Figure~\ref{figure:logIngestionExample}.
The ingestion optimizer takes this DAG and emits an optimized ingestion plan. To do so, the optimizer supports rule-based tree transformations to identify subtree patterns and transform them into alternate subtrees.
An ingestion plan subtree is represented as an ingestion operator expression, which consists of the root operator and its descendants (recursively).
An optimizer rule (for tree transformations) operates on the ingestion operator expression via the following two methods:

\vspace{-0.2cm}
\begin{small}
$\texttt{check}: \textit{IngestOpExpr} \rightarrow \textit{true/false}$\\
\indent $\texttt{apply}: \textit{IngestOpExpr} \rightarrow \textit{IngestOpExpr}'$
\end{small}
\vspace{-0.1cm}

\noindent The \texttt{check} method verifies whether a rule is applicable to an ingestion operator expression and the \texttt{apply} method produces the modified ingestion operator expression.
The optimizer performs a preorder traversal over the ingestion DAG and fires matching rules wherever applicable, i.e., larger subtrees are matched for relevant rules first.
The rules are matched in the same sequence as provided in the ordered rule set, and they are applied iteratively until none of the rules match any of the ingestion operator expressions in the tree.
We now describe two rules, namely operator reordering and pipelining, to reduce the data volume and the materialization cost respectively.


\noindent\textbf{Operator Reordering.} This rule rearranges ingestion operators in order to reduce the data volume in flight, i.e.,~push-down data reducing operators, e.g., filter, while push-up the data expanding operators, e.g., replicate.
In order to preserve the semantics of the ingestion plan, we only rearrange the ingestion operators in the same data flow stage, i.e., there is no conditional processing of the data items involved.
One instance of this rule could push replicate operator at the very end of the stage, i.e.,~replicate data as late as possible, as shown in stage $b$ of Figure~\ref{figure:logIngestionExampleReordering}.
Another instance could swap the filter and projection operator depending on which provides more data reduction, i.e.,~whether we reduce data volume more by filtering the rows or by filtering the columns.
Thus, operator reordering rules could be useful in reducing the data traffic while executing the ingestion plans.



\noindent\textbf{Operator Pipelining.} By default, all output ingest data items are collected (i.e., materialized) from an ingest operator before being fed to the next one.
Internally, this is done by adding a \textit{materialize} operator after each ingest operator.
An obvious optimization is to pipeline the data items between operators as much as possible and materialize only when really needed.
Operator pipelining rule removes materialization between operators that process ingest data items of the same granularity (detected by looking at the data types).
We materialize only when the granularity of the ingest data item changes, e.g.,~from tuples to blocks.
To illustrate, Figure~\ref{figure:logIngestionExamplePipelining} shows the log ingestion plan from Section~\ref{sec:logIngestionExample} with operator pipelining.
We can see that stages $a-g$ of the plan, as shown in Figure~\ref{figure:logIngestionExample}, have been transformed to five pipelined blocks $1-5$ in Figure~\ref{figure:logIngestionExamplePipelining}.
Other instances of the operator pipelining rules could consider materializing long pipelines in between for fault-tolerance, or for early access to the incoming data.

Thus, we see that the ingestion optimizer provides an extensible way to transform and optimize the ingestion DAGs.

\section{\system{} Runtime Engine}
\label{section:execute}


Recall that the modern big data applications need to ingest the incoming data quickly and with low overhead.
As a result, it is critical to have an efficient runtime engine for these applications.
In this section, we describe the \system{} runtime engine which (i)~runs an ingestion plan in parallel on a cluster of machines,
(ii)~efficiently handles distributed data I/O during ingestion, and
(iii)~handles fault-tolerance both during and after ingestion.
We describe each of these below.

%
%


\subsection{Parallel Ingestion}
\label{sec:parallel}

Given an ingestion plan and a cluster of machines, \system{} runtime engine exploits two kinds of parallelism: {\it inter-node} and {\it intra-node parallelism}. We describe these below.

\noindent\textbf{Inter-node Parallelism.} 
When a user submits an ingestion plan on one of the nodes (the client) for execution, the \system{} runtime engine copies the resulting optimized plan to all nodes (specified via a \textit{slaves} configuration file) in the cluster and executes it over the local data on each node. This makes sense because the raw data is typically generated on multiple nodes in the first place, e.g.,~log data, and it is cumbersome to bring all of this data to a single node. Therefore, instead of bringing data to the ingestion plan, we ship the plan to the data itself. This is similar to shipping query plans in distributed query processing.
\system{} runtime engine launches remote shell to start the ingestion plans on all nodes in parallel and waits for them to finish before it terminates.


\noindent\textbf{Intra-node Parallelism.} Besides parallelizing the ingestion process across different nodes, the \system{} runtime engine also parallelizes part of the ingestion plan across different threads on the same node. For example, the \texttt{serialize} operator is CPU bound and so the \system{} runtime engine forks several operator instances (as many as the number of cores by default) at the same time, each serializing a different subset of ingest data items. Likewise, the \system{} runtime engine transforms different replicas of ingest data items in different threads. 
To support such multi-threaded parallelism, the ingest operator implementation has a parallel mode, in addition to the default serial mode, to process input ingest data items using a thread pool.
These threads are later synchronized in the \textit{finalize} method of the ingestion operator.
The parallel mode is turned on by default for CPU heavy operators such as \texttt{serialize}. However, users could provide additional optimizer rules to control the serial/parallel modes.


Parallel ingestion allows \system{} to significantly reduce the overhead of transforming the data. We demonstrate this experimentally in Section~\ref{section:experiments}.

\subsection{Efficient Distributed I/O}
\label{subsection:datamovement}

In the previous section, we described how we can parallelize the ingestion plan and process data locally on each node.
However, several ingestion plans require to move data around.
In this section, we describe how the \system{} runtimes engine handles distributed I/O efficiently.
Below we describe the three major data movement scenarios, namely \textit{shuffling}, \textit{placement}, and \textit{replication}.

\noindent\textbf{Shuffling.} An ingestion plan may require to shuffle intermediate ingest data items in order to produce the final data items.
For example, to gather stratified samples, we need to group the entire dataset across all nodes and then pick samples from each group\footnote{Users could also do per-node stratified sampling to compute the samples from the local stratum on each node.}.
\system{} runtime engine handles this using a distributed file system by first creating local groups on each node and then copying them to the distributed file system in parallel.
While copying, the data is organized into directories, one for each group, such that data belonging to the same group is in the same directory.
Finally, each node reads back and processes the group-directories, one at a time, from the distributed file system.
Essentially, the \system{} runtime engine leverages the remote data access mechanism of the distributed file system to shuffle data across nodes.

\noindent\textbf{Placement.} \system allows users to reason data placement at a logical level, i.e.,~using the \textit{locator} operator to map each ingest data item to a location ID, without getting into the low level data placement policies. As a result, users can easily make data placement decisions, such as which portions of the data should reside on which nodes; or, which data items should be co-located and which data items should \textit{not} be co-located. To enforce these decisions, the \system{} runtime engine simply looks at the location IDs of each ingest data item, e.g., a data block, and copies items with the same location ID to the same node in the cluster. The mapping from location IDs to nodes can either be provided by the user, or the runtime self-assigns the location IDs to nodes in the same order (in a round robin manner) as they appear in the \textit{slaves} file.



\noindent\textbf{Replication.} Replication is usually done for fault-tolerance and it typically involves moving each replica to a different node.
In contrast, \system completely decouples data replication and placement, and allows users to take independent decisions for the two.
As a result, users can choose to replicate data at different granularities and/or may not place the replicas on different nodes.
For example, users may choose to replicate some rows (could be seen as samples) in each data block and store them along with the data block on the same node, i.e.,~no additional data movement is needed.


\subsection{Fault Tolerance}
\label{subsection:faulttolerance}

In this section, we describe the fault-tolerance mechanisms in \system{} runtime engine to handle failures both during and after the data ingestion.

\subsubsection{Handling In-flight Failures}

The \system runtime engine can handle two types of failures while running the ingestion plan.

\noindent\textbf{Ingestion Operator Failure.} In case an ingestion operator fails, we need to re-run all pipelined operators that appear before the failed one.
However, instead of restarting the ingestion plan from scratch, we can resume ingestion from the previous block of pipelined operators.
This is because ingest data items are fully materialized after every pipelined operator block, and therefore each such block serves as a checkpoint.
In case of repeated failures (3 times by default) of the same operator\footnote{The \system{} runtime engine detects recurring failures by keeping track of the execution status of each ingestion operator, i.e., whether or not they passed the \textit{finalize} method successfully.}, the \system{} runtime engine replaces the failing operator with a dummy pass-through operator.
The dummy operator simply returns the input ingest data items and assign each item a label of ``$-1$" to denote the failure.
Application developers can further control the recovery time (e.g., in case they expect more failures) by adding custom operator pipelining rules to force more frequent materialization, as discussed in Section~\ref{section:optimizer}.

\noindent\textbf{Node Failure.} In case one of the nodes in the cluster fails, we simply reschedule the ingestion plan from the failing nodes to other nodes. 
However, this still requires the data on the failed node to be available remotely (in case that node is used as the data node as well).
For node failures during data shuffling, we check which of the groups directories in the distributed file system are corrupt and we copy them again, assuming that the distributed file system still works with one less node.
To handle data placement, we reassign the location ID of the failed node to the next node (in \textit{slaves} file) in the round robin sequence.

\subsubsection{Handling Post-ingestion Failures}


Given that data is ingested with custom application-specific logic, it may also need custom fault-tolerance logic.
\system{} allows users to control the fault-tolerance mechanism for their data based on their ingestion plans.
To do so, \system{} provides two fault-tolerance UDFs to define how to detect and recover failing data items (typically data blocks).

\vspace{-0.2cm}
\begin{small}
$\texttt{detect}: f \rightarrow \{r_1,r_2,..,r_n\}$\\
\indent $\texttt{recover}: \{B_{r_1},B_{r_2},..,B_{r_n}\} \rightarrow B_f$
\end{small}

\noindent Here $f$ is the failing block id while $\{r_1,r_2,..,r_n\}$ are the recovery blocks; $B_i$ is the corresponding block. The above UDFs essentially address two key questions: (1)~Which data blocks are needed to recover a failed data block? (2)~How to reconstruct a failed data block from the recovery data blocks?

\begin{figure}[!t]
\hspace{-0.1cm}
\includegraphics[width=3.3in]{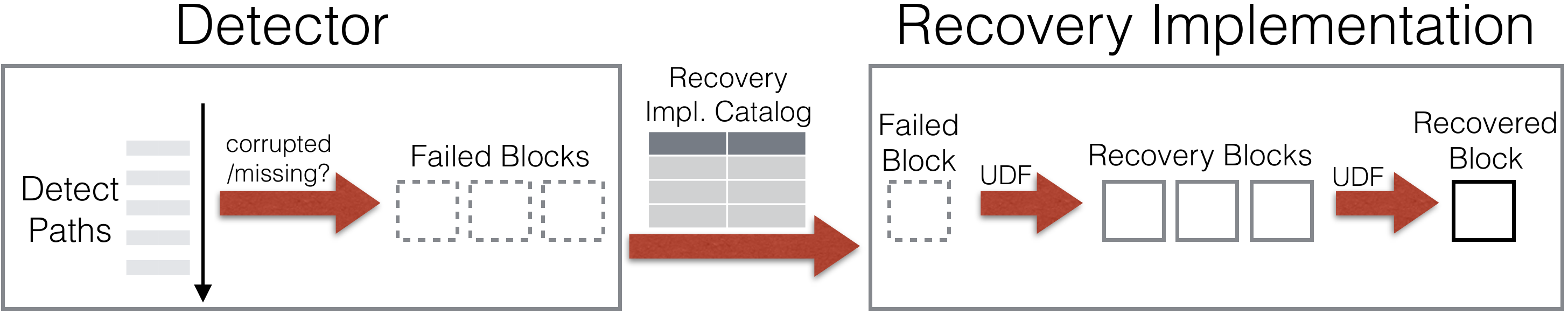}
\vspace{-0.1cm}
\caption{Post-ingestion fault-tolerance in \system.}
\vspace{-0.5cm}
\label{figure:ft_arch}
\end{figure}

As soon as it finishes executing the ingestion plan, the \system{} runtime engine 
launches a fault-tolerance daemon that polls the storage system for failing data blocks (the \texttt{detect}) and invokes the user-supplied recovery UDFs for every failed block detected (the \texttt{recover}). 
\system{} maintains a catalog of \texttt{detect} and \texttt{recover} UDFs for each ingestion plan.
Figure~\ref{figure:ft_arch} depicts this post-ingestion fault-tolerance mechanism in \system{}.

We built three implementations of the detect and recover UDFs using the above architecture:

\noindent\textbf{Replication based.} This fault-tolerance mechanism looks for a replica of the failed data block and increases the replication factor of the replica by 1. The block placement policy takes care of storing the new replica on a different node.

\noindent\textbf{Transformation based.} This recovery mechanism is for data block replicas that are not bitwise identical, i.e.,~they are serialized differently. This mechanism copies and transforms a data block replica so that it has the same serialization as the failed data block.

\noindent\textbf{Erasure coding based.} This recovery mechanism is for erasure-coded, instead of replicated, data blocks. It first fetches all data blocks in the same stripe and then reconstructs the missing data block. The reconstructed data block is stored back to the HDFS. 

Thus, we see that \system{} users can: (i)~inject custom fault-tolerance logic for their application specific needs, e.g.,~heterogeneous replication~\cite{hail,trojanlayout}, (ii)~change the fault-tolerance over time as the application needs evolve, e.g.,~migrating from replication to erasure coding~\cite{hdfs-raid}, and (iii)~have different fault-tolerance mechanisms for different ingestion plans, i.e.,~fault-tolerance mechanism is not tied to the storage system anymore. 
\section{Ingestion Aware Data Access}
\label{section:dataaccess}


\system allows users to apply ad-hoc data transformations while ingesting their datasets. However, the system also needs to keep track of these transformations in order to leverage them later for query processing. Essentially, we need to track three pieces of information: (i)~which ingestion operators were used to preprocess the dataset; (ii)~how were the ingestion operators composed; and (iii)~the operator lineage and the transformation applied to each output data item. For (i) and (ii), we simply serialize the ingestion plan in the storage system. Note that we do not serialize the operator instances, rather we store the instance parameters and re-instantiate the operators whenever needed. For (iii), we make use of labels assigned to the ingest data items, as described below.

Recall that each ingestion operator assigns a label to every data item that is processed.
One could imagine storing all such labels for every ingest data item. However, this would result in a huge amount of metadata. Instead, the \system{} runtime engine collects the labels common to all data items that are materialized and preserves them as the name of the physical file. 
Thus, each physical file in \system is named as follows: $\textit{label1\_label2\_label3\_label4\_.....labeln}$.
The labels in the above filename have the same relative sequence as the corresponding operators in the ingestion plan. Thus, the filename of a physical file in \system acts as a signature, or the {\it lineage} of the preprocessing applied to it. For example, the name of a physical file produced by the ingestion plan of log analytics (Figure~\ref{figure:logIngestionExample}) might be: \textit{\small parseID\_replicaID\_hashID\_fileID\_paxID\_locationID\_uploadID}.
As a result of these label encoded filenames, \system does not need to maintain any additional metadata files.

Once the data is ingested using \system{}, users want to access it from their applications.
\system{} provides ingestion-aware access methods to query data from arbitrary query processors.
Again, the \system{} access methods address three key questions: (i)~\textit{what} to access, (ii)~\textit{where} to access, and (iii)~\textit{how} to access.
We describe these three below.

\noindent \textbf{What to access?} \system{} allows developers to retrieve a subset of a dataset, based on the labels applied to the ingested data items. To do so, \system{} provides two filter operators:  one that filters data replicas and one that filters data blocks in a particular replica:

\vspace{-0.5cm}
\begin{small}
\begin{align*}
& \textit{filterReplica }\text{(\texttt{\scriptsize IngestOp} filterOperator, \texttt{\scriptsize Label} operatorLabel)} & \\[-0.3em]
& \textit{filterBlock }\text{(\texttt{\scriptsize IngestOp} filterOperator, \texttt{\scriptsize Label} operatorLabel)} &
\end{align*}
\end{small}
\vspace{-0.6cm}

\noindent As an example, consider a sampling ingest operator that labels every data item as either $1$ (denoting a sample) or $0$ (denoting the original data item). Also assume that the ingestion plan physically partitions the sampled and original ingest data items into different physical files. To access only the samples, we can use {\tt\scriptsize filterReplica} to  filter the files that have label $1$. This narrows down data access only to the relevant portions of data.

\noindent\textbf{Where to access?} In addition to filtering, \system{} allows developers to define: (i)~the data access parallelism by setting the number of tasks to run in parallel; and (ii)~the amount of data each computation task has to read. This is done by assigning data blocks to computation tasks. \system API allows key-based splitting as well as co-splitting two or more datasets on their respective keys. 

\vspace{-0.5cm}
\begin{small}
\begin{align*}
& \textit{splitByKey }\text{(\texttt{\scriptsize Key} key [, \texttt{\scriptsize Int} maxSplitSize])} & \\[-0.3em]
& \textit{coSplitByKey }\text{(\texttt{\scriptsize Key} key1, \texttt{\scriptsize Dataset} d2, \texttt{\scriptsize Key} key2, ..)} &
\end{align*}
\end{small}
\vspace{-0.6cm}

\noindent For example, if the ingestion plan partitioned the data on an attribute into ranges, developers can distribute different range partitions to different machines and increase data access parallelism; or to the same machines to improve data locality.

\noindent\textbf{How to access?} Finally, \system{} allows developers to deserialize the retrieved blocks and apply further selection/projection predicates while reading them.

\vspace{-0.6cm}
\begin{small}
\begin{align*}
& \textit{deserialize}\text{(\texttt{\scriptsize Projection} p, \texttt{\scriptsize Selection} s)} &
\end{align*}
\end{small}
\vspace{-0.6cm}

Note that the actual deserialization depends on the serialization operator in the ingestion plan.  The built in \system library provides deserialize operators for all of the serialize operators it provides (PAX, RCFile, SortedFile, ColumnGroup, etc.)
These implementations take into account the selection/projection predicates while deserializing the data. For example, they may deserialize only the projected attributes (in case of column layout), or perform index access (in case the data is sorted).

Ingestion-aware data access pushes down one or more query predicates before producing the input for the upstream query processor.
The following section describes how one could use these access methods with two popular query processing engines, namely Hadoop MapReduce and Spark.

\section{Integrating \system{}}
\label{section:prototype}

We now describe how \system{} works with two different combination of storage and compute substrates. 

\begin{figure}[!t]	
\hspace{-0.2cm}
\includegraphics[width=3.4in]{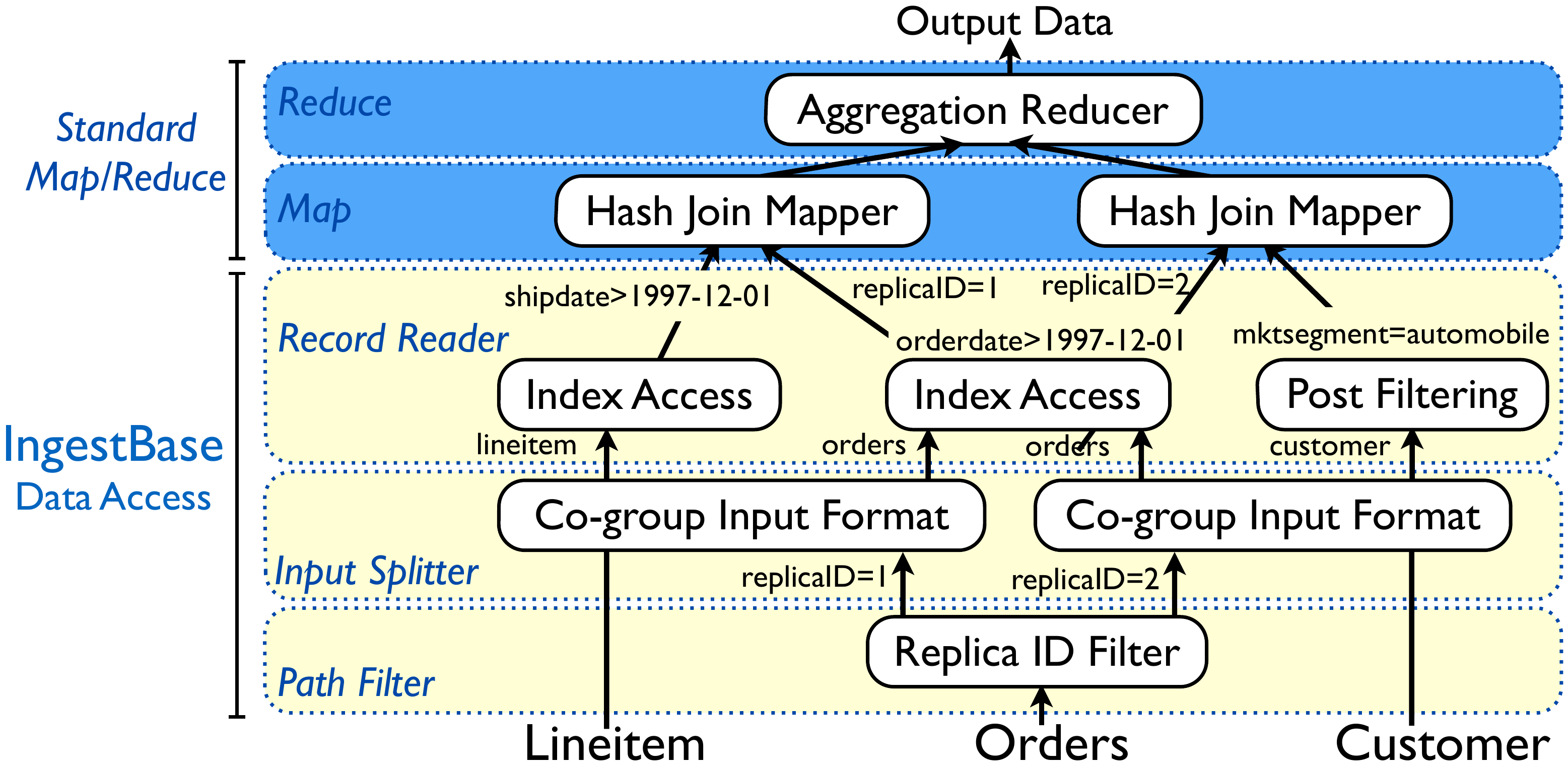}
\caption{Ingestion-aware data access for TPC-H Q3 using Hadoop MapReduce.}
\label{figure:queryplan_q3}
\vspace{-0.2cm}
\end{figure}

\subsection{HDFS \& MapReduce}

Let us first look at how \system interacts with HDFS.
First of all, \system needs to map the ingest data items to physical HDFS files, i.e.,~collect the output ingest data items from an ingestion plan and store them in HDFS. To do so, the last transformation in an ingestion plan must be \textit{upload}. If the ingestion plan contains a physical partitioner, the \textit{upload} operator maps each physical partition to a HDFS file. Otherwise, it collects all data items into a single HDFS file. \system further controls several storage decisions of the HDFS files it creates. For instance, it can replicate each physical file if the ingestion plan already contains a replication operator, split files into subfiles and choose a different replication for each subfile, or let HDFS do the standard 3x replication. Likewise, \system controls data placement by assigning location IDs to physical partitions and mapping each location ID to a particular data node, via a custom data placement policy. Similarly, the \textit{upload} could pipeline the data items produced by a ingestion plan directly to HDFS files, without first collecting them on local disk. It can also bulk load the data items to HDFS files. Finally, \system can manipulate the fault-tolerance mechanism, e.g.,~transform data layouts when recovering failed blocks (Section~\ref{subsection:faulttolerance}).
Thus, even though \system sits on top of HDFS, it could be tightly integrated with the storage decisions in HDFS.

For MapReduce query processing, we bake \system{} access methods using the \textit{Hadoop InputFormats}. 
The InputFormat allows users to specify a \textit{path filter} to filter input based on the HDFS file path, a \textit{splitter} to split the data logically, and a \textit{record reader} to actually read the data.
We implemented custom functionality for these three methods in order to use the \system access methods in Hadoop.
For example, to implement the \textit{filterReplica}, we created a path filter which retrieves all physical files having a particular label in their filename\footnote{Recall that we persist the labels in the filenames.}.
We also implemented additional helper methods, e.g.~\textit{filterReplicaById}, \textit{filterReplicaByPartitioning}, and \textit{filterReplicaByLayout}, for the ease of programming. 

To illustrate, Figure~\ref{figure:queryplan_q3} shows how the \system access methods can be used to run TPC-H Q3 (which consists of two joins and a GROUP BY) in a single MapReduce job, in contrast to two jobs in standard Hadoop. This is possible because \system co-groups all three TPC-H relations.
Note that the output of \system{} access methods is fed to standard map/reduce data flows.
Thus, in addition to allowing users to easily preprocess and transform their datasets, \system access methods also allows developers to quickly build efficient query processors.

\begin{figure*}[!t]	
\hspace{-0.3cm}
\subfigure[Data Cleaning]{
\includegraphics[scale=0.42]{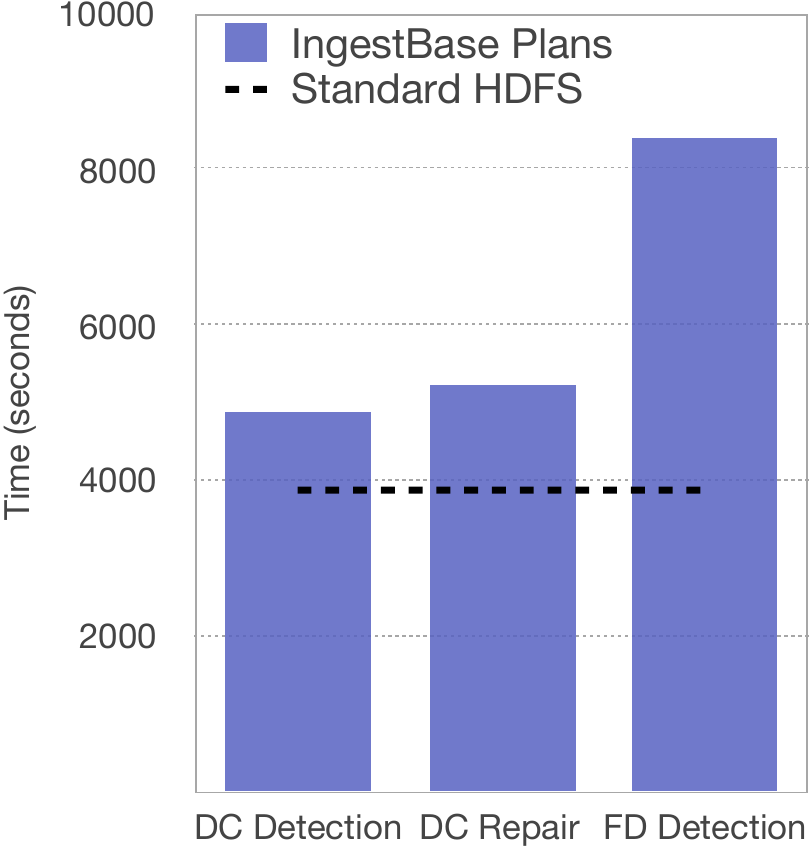}
\label{figure:cleaning}
}
\hspace{-0.25cm}
\subfigure[Data Sampling]{
\includegraphics[scale=0.42]{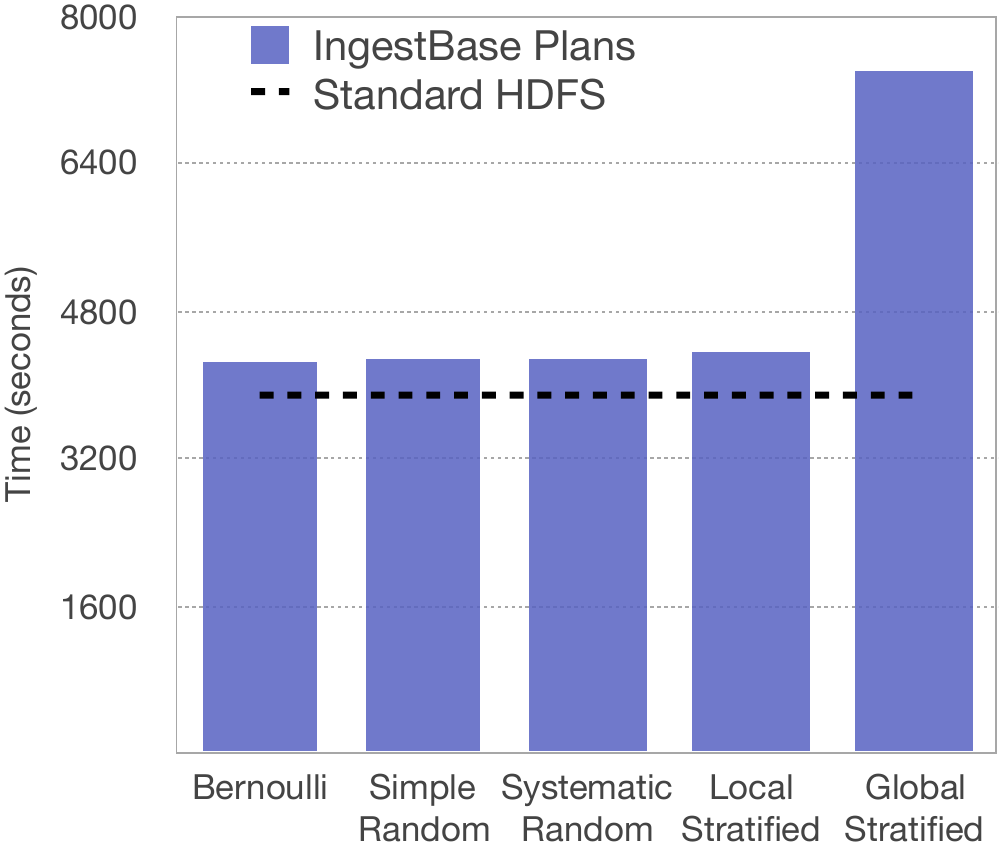}
\label{figure:sampling}
}
\hspace{-0.25cm}
\subfigure[Data Analytics]{
\includegraphics[scale=0.42]{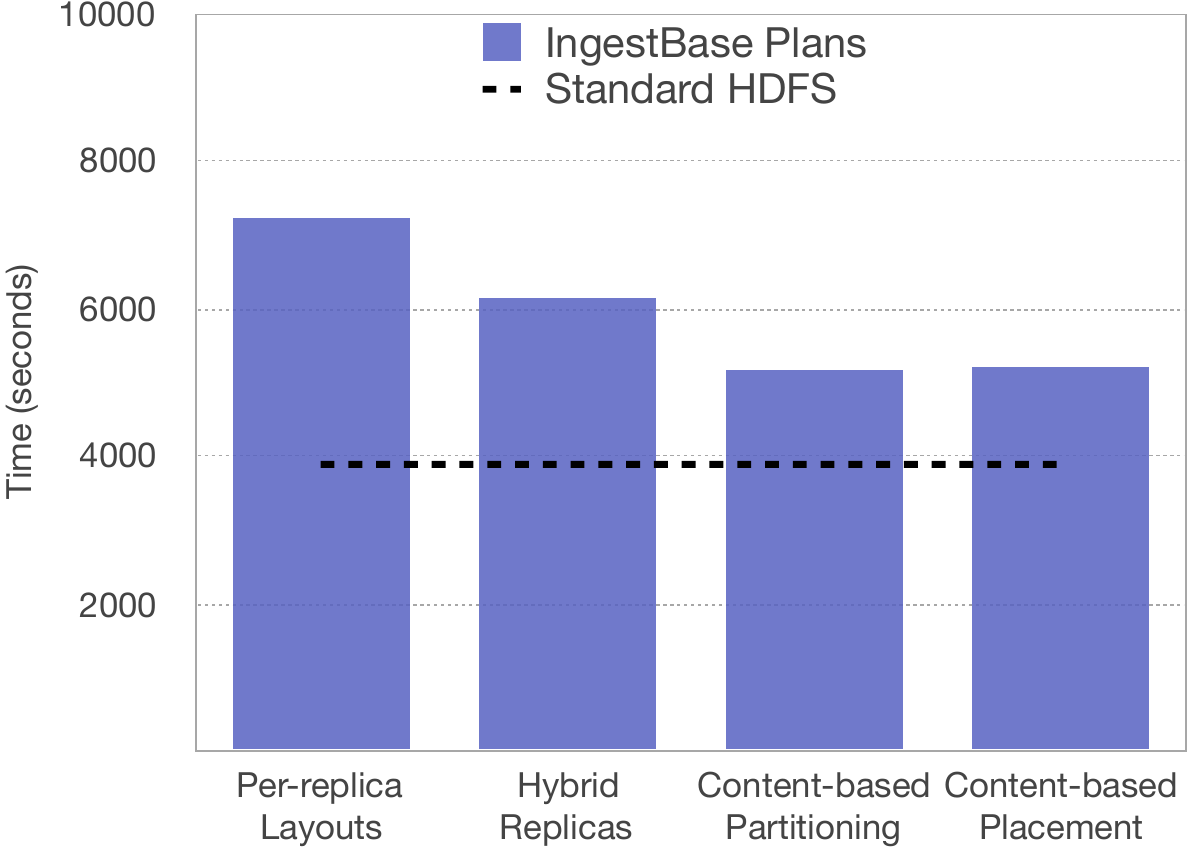}
\label{figure:analytics}
}
\hspace{-0.25cm}
\subfigure[Data Storage]{
\includegraphics[scale=0.42]{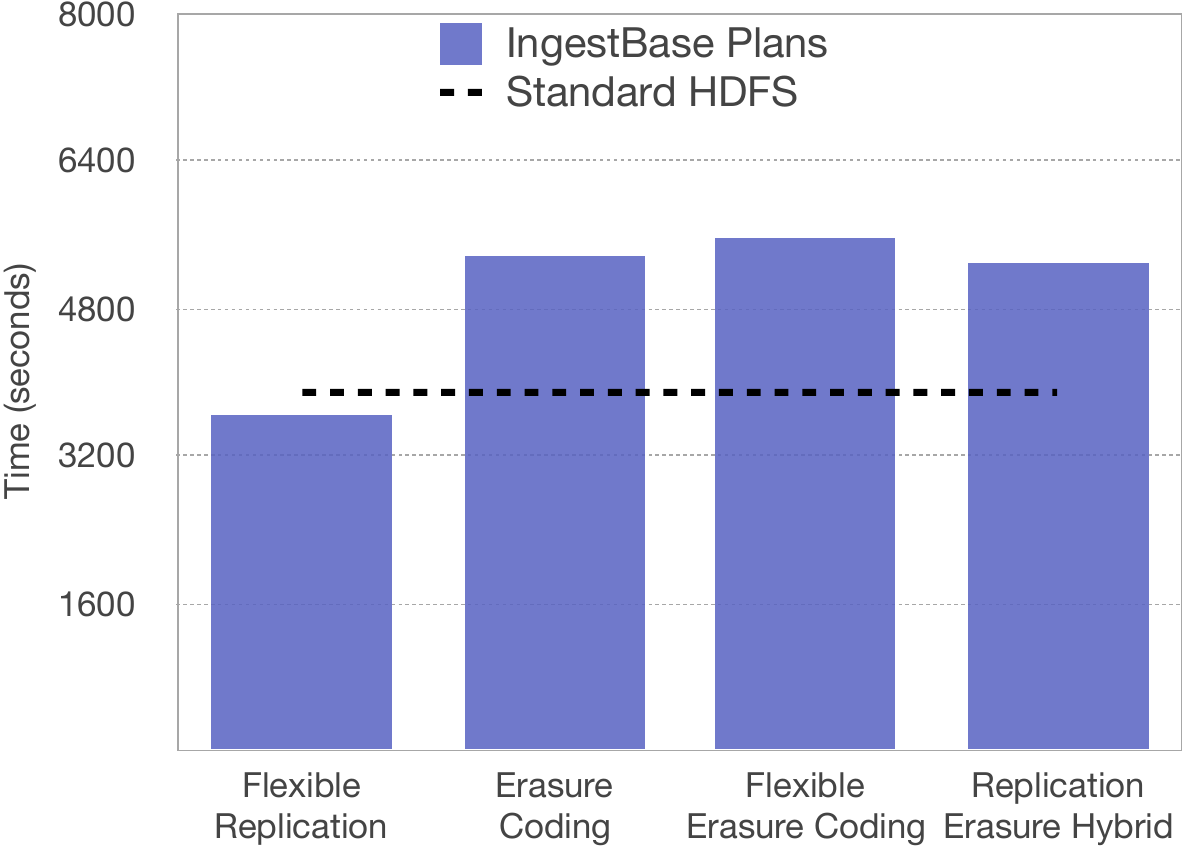}
\label{figure:storage}
}
\vspace{-0.3cm}
\caption{Ingestion runtime engine overhead of \system{} over different applications, compared to uploading to standard HDFS (without any preprocessing).}
\label{figure:ingestionPerformance}
\vspace{-0.4cm}
\end{figure*}

\subsection{HDFS \& Spark}

Spark runs over HDFS as well as it uses the same \textit{Hadoop InputFormats} to read data from HDFS.
As a result, we could easily run Spark jobs on top of \system{} ingested data.
To illustrate, Listing~\ref{listing:sparkaccess} shows the data access for group-wise analytics over sampled data in Spark.
This data access plan selects the replica of the sampled data which is partitioned on {\tt\small PARTKEY}, co-locates (splits) values of {\tt\small PARTKEY}, and projects {\tt\small PARTKEY} and {\tt\small SUPPKEY} attributes. Finally, we get an RDD from the ingested dataset and can apply standard Spark transformation over it. We see that using the \system data access plans, developers can easily narrow down their analysis to the most relevant portions of the data, without dealing with the actual physical data representation used to store the data.

\begin{lstlisting}[basicstyle=\scriptsize\normalfont\sffamily,aboveskip=0pt,belowskip=-6pt,float=!t,language=java,label=listing:sparkaccess,caption=Ingestion-aware data access for group-wise analytics using Spark.]
public JavaRDDLike<?,?> groupwiseAnalytics(String ingestFilepath) {
  JavaSparkContext ctx = new JavaSparkContext(SPARK_MASTER, ``myJob", SPARK_HOME,SPARK_JAR);

  // IngestBase data access
  IngestBaseDataset d = new IngestBaseDataset(ctx, ingestFilepath);    
  d.filterBlock(SamplingOperator, SAMPLE_ID);
  d.filterReplicaByPartitioning(PARTKEY);
  d.splitByPartitionKey(PARTKEY);   
  d.deserializeProject(PARTKEY, SUPPKEY);
  
  // standard Spark transformations
  return d.RDD()
          .map(new GroupbyKeyMap())
          .reduceByKey(new GroupbyKeyReduce());
}
\end{lstlisting}

\noindent

\hide{
\subsection{Other Storage \& Compute Substrates}

The \system{} design is general enough to also work with other storage and compute engines, other than Hadoop and Spark.
For instance, other HDFS-like file systems and distributed batch query processors, e.g., Cosmos/SCOPE~\cite{cosmosScope}, could easily integrate with \system{}.
Even traditional database users increasingly ingest data for a variety of newer applications.
Examples include, ingesting graph data for graph analytics in a column store~\cite{graphsOnVertica}, ingesting operational analytics data in a row store~\cite{columnIndexes}, and ingesting machine learning data into a database~\cite{madlib}.
Each of these applications process data differently and would therefore like to ingest data differently, thus making the \system{} design applicable.
}
\section{Experiments}
\label{section:experiments}

We ran a number of experiments to evaluate \system.  Our goal was to answer two key questions:
(i)~{\it how efficiently does \system allow users to perform data transformations?} and
(ii)~{\it is transform-as-you-upload in \system better than other possible transformation approaches?}
To evaluate these questions, we ran \system ingestion plans for the four different data ingestion scenarios described in Section~\ref{section:ingestionpain}. For all  experiments, we measure unmodified HDFS data upload times as the default baseline, and Hive (a widely used SQL-based database that runs on Hadoop MapReduce and HDFS) as an additional baseline wherever possible.
Additionally, we evaluate the data access times \system for several common relational operations. All experiments were done on a cluster of $10$ nodes. 
Each node has $1.07$ GHz with 32-core Xeon running on Ubuntu 12.04, $256$ GB main-memory, and $11$ TB of disk storage. We experimented with TPC-H data at scale factor $1000$ ($1$TB in total), and generated data on all $10$ nodes in parallel. We run \system on top of Hadoop 2.0.6-alpha and use Hive 0.13.1.

\subsection{Data Ingestion Scenarios}
In this section, we describe the performance of \system on the four ingestion scenarios described in Section~\ref{section:ingestionpain}.

\subsubsection{Data Cleaning}
\label{sec:cleaning}
We start by evaluating \system on data cleaning operations, when uploading the TPC-H \texttt{\small lineitem} table into HDFS. 

\noindent{\bf Setup.} We consider the data quality rules described in Section~\ref{section:cleaning}: (i)~a {\it functional dependency} (FD) stating that any two tuples having the same \texttt{\small ship\_date} must have the same \texttt{\small line\_status}; and (ii)~a {\it denial constraint} (DC) stating that any tuple having \texttt{\small quantity} smaller than $3$ must have \texttt{\small discount} smaller than 9\%.  
We measure the runtime of three different transformations: (i)~detecting FD violations, (ii)~detecting DC violations, and (iii)~repairing (in addition to detecting) DC violations.

\noindent{\bf Discussion.} Figure~\ref{figure:cleaning} shows the results. We observe that detecting violations for the DC rule while ingesting data incurs only a $25\%$ overhead over standard HDFS. This is because the detection process is simply piggy-backed into the process of uploading data into main-memory. This is also why DC repair incurs almost no extra overhead. However, this overhead increases when the data quality rules require more complex data transformations. 
For instance, the ingestion plan to detect FD violations takes double the standard HDFS upload time.
This is because the FD requires grouping the entire dataset on \texttt{\small ship\_date}, which results in shuffling the data across all nodes.
Still, detecting violations when ingesting datasets (using \system) is much better than detecting violations after a dataset is ingested (using e.g.,~Hive), as we shall see in Section~\ref{subsubsec:hive_baseline}.

\subsubsection{Data Sampling}

We  now look at the performance of \system for computing samples during data ingest.

\noindent{\bf Setup.} We consider five different sampling techniques: {\it Bernoulli}, {\it simple random}, {\it systematic random}, {\it local stratified}, and {\it global stratified}. Here the local stratified sampling collects samples from the local strata on each node, whereas the global stratified sampling collects samples from the global strata.

\noindent{\bf Discussion.} Figure~\ref{figure:sampling} shows the results of these experiments. We observe that \system has a very small overhead (less than $10\%$) for all methods except global stratified sampling.
This small overhead reflects the time to write the data samples to disk. In the case of global stratified sampling,  \system upload time is nearly twice of HDFS upload time. The reason is the same as for the DC rule in the previous experiments: global stratified sampling  requires shuffling the entire dataset across all nodes to collect samples from each subgroup.
However, these experiments show that most types of sampling can be done  efficiently, as data is being ingested, with little additional overhead (no additional passes over the entire data set).

\subsubsection{Data Analytics}

In this section, we analyze the \system ingest times when preparing datasets for different data analytical tasks. 

\noindent{\bf Setup.}  We create different layouts for each replica (using the scheme the Trojan Layouts~\cite{trojanlayout} paper).  We denote this scheme as {\it Per-replica Layouts}.
It works by creating a binary row, PAX, and compressed PAX layouts for each of the three replicas, respectively.
In addition, we implement three new data storage schemes, as described in Section~\ref{subsection:dataanalytics}: (i)~{\it Hybrid Replicas}, which store subsets of the blocks of a replica in different layouts, (ii)~{\it Content-based Partitioning}, which chunks the data based on its content instead of physical size, and (iii)~{\it Content-based Placement}, which places data blocks based on their content. 
For Hybrid Replicas, we create the same three layouts as in Per-replica Layouts and we let HDFS handle the replication.
For the other two schemes, we use a logical partitioner to generate $10$ range partitions (for the Content-based Partitioner) and place all those data blocks having the same range on the same data node (for the Content-based Placement).

\noindent{\bf Discussion.} Figure~\ref{figure:analytics} illustrates the results, which overall confirm the trend we observed in the previous section: the overhead is directly proportional to the time spent by \system in transferring or processing data. We observe that when creating a different data layout per data replica (Per-replica Layouts), the \system ingest time is nearly double the HDFS upload time. This is mainly because \system has to deal with data replication outside HDFS. When \system pushes data replication to HDFS itself (such as in Hybrid Replicas, Content-based Partitioning, and Content-based Placement), we  see a decrease of the \system overhead. In particular, we observe that for the Content-based Partitioning and Content-based Placement schemes, which are less CPU demanding than the other two schemes, the overhead decreases even more, to just over 20\%. These results show the efficiency of \system at applying arbitrary data transformations at ingest time, and without any modifications to the HDFS.

\subsubsection{Storage Space Optimization}

Finally, we evaluate \system in scenarios where optimizing the storage space is the primary goal of the user. 

\noindent{\bf Setup.} We consider four different scenarios. First, the case where users over-replicate the hot data, for better availability, and under-replicate the cold data, for preserving storage space (Flexible Replication). To do so, we create $10$ range partitions and consider the first partition to be hot  (replicated $10$ times) and the remaining partitions to be cold  (replicated $2$ times). Second, we consider using erasure coding instead of replication. We erasure code the data with $3$ parity blocks for every $10$ data blocks. Third, we consider the case where users apply different erasure codes to different portions of their data (Flexible Erasure Coding): we create $3$ parity blocks for every $5$ data blocks of the first range partition, while the remaining range partitions are encoded as before, i.e.~$3$ parity blocks for every $10$ data blocks. Finally, we consider the case where users apply both replication and erasure coding, on different portions of the data: we replicate the first range partition $10$ times and apply erasure coding for the remaining partitions ($3$ parity blocks for every $10$ data blocks).

\noindent{\bf Discussion.} Figure~\ref{figure:storage} shows the results. Interestingly, we observe that \system outperforms HDFS in the Flexible Replication case. This is because \system creates fewer data replicas than HDFS and hence  stores less data. 
Erasure coding, on the other hand, stores $30\%$ more data as well as incurs more CPU costs. As a result, it is has almost $40\%$ higher runtime than standard HDFS. 
However, \system allows developers to flexibly choose the erasure codes as well as freely combine erasure coding with replication.
Hence, \system can be effectively used to optimize storage space by transforming the physical data representation in a variety of ways.

\begin{figure*}[!t]
\hspace{-0.2cm}
\subfigure[Projection]{
\includegraphics[scale=0.42]{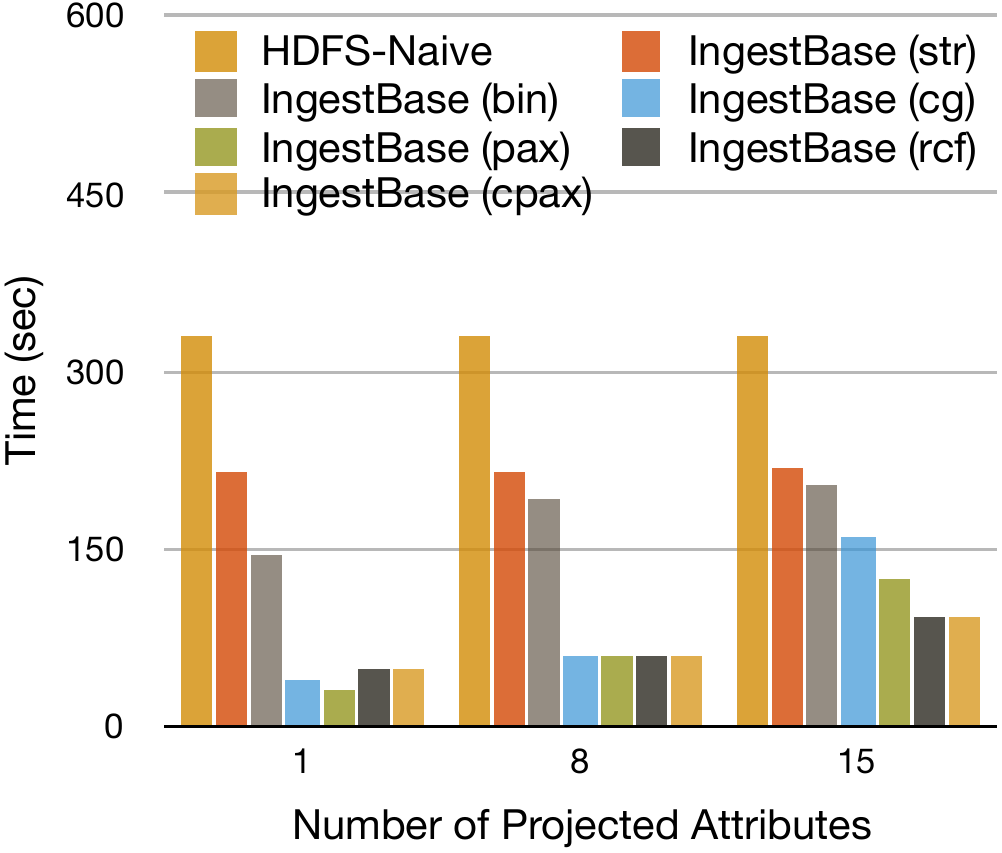}
\label{figure:operator_projections}
}
\hspace{-0.25cm}
\subfigure[Selection]{
\includegraphics[scale=0.42]{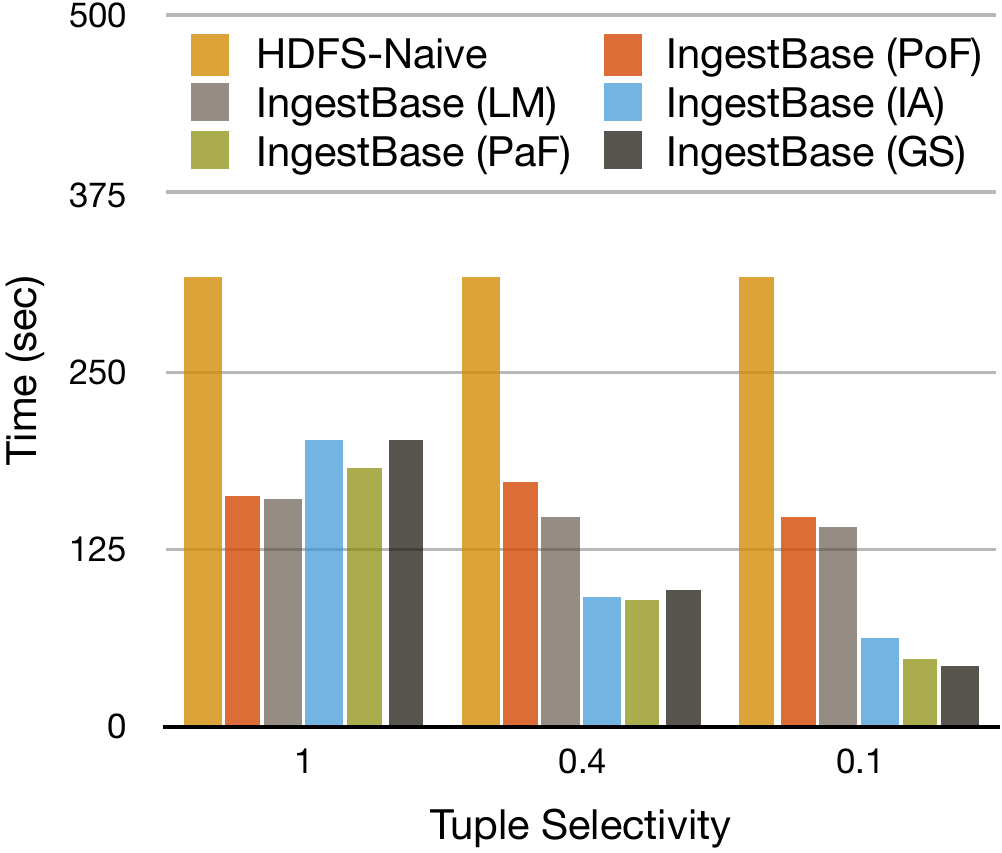}
\label{figure:operator_selections}
}
\hspace{-0.25cm}
\subfigure[Aggregation]{
\includegraphics[scale=0.42]{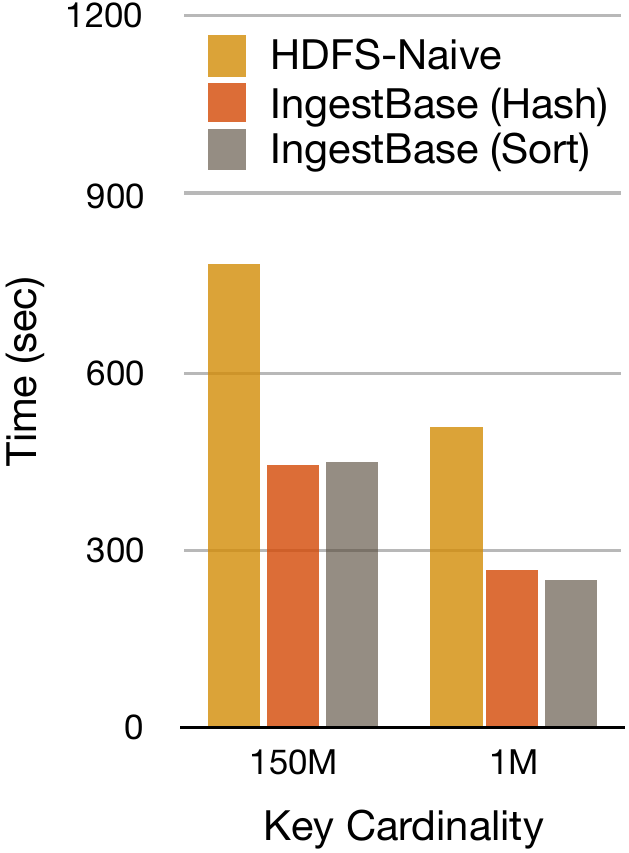}
\label{figure:operator_aggregation}
}
\hspace{-0.25cm}
\subfigure[Join]{
\includegraphics[scale=0.42]{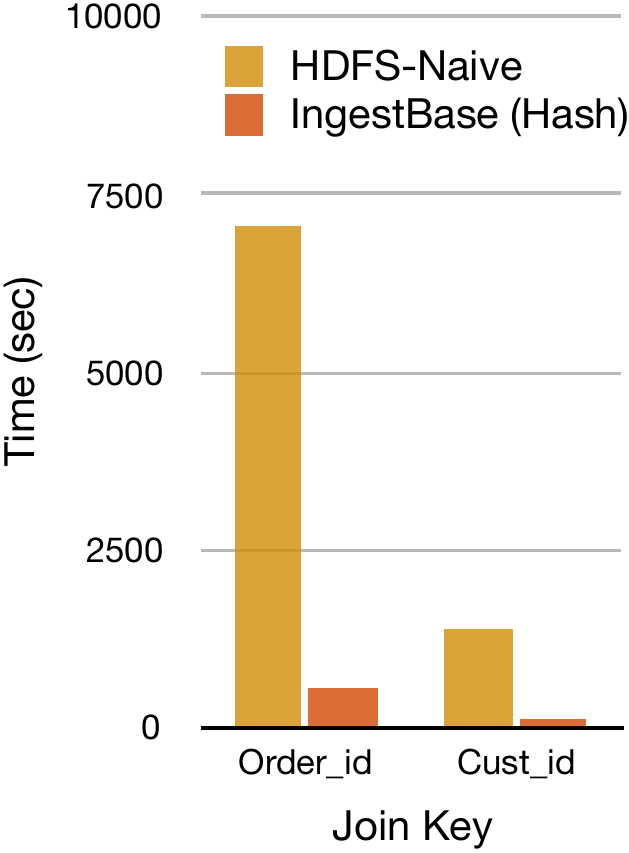}
\label{figure:operator_joins}
}
\hspace{-0.25cm}
\subfigure[Full Query]{
\includegraphics[scale=0.42]{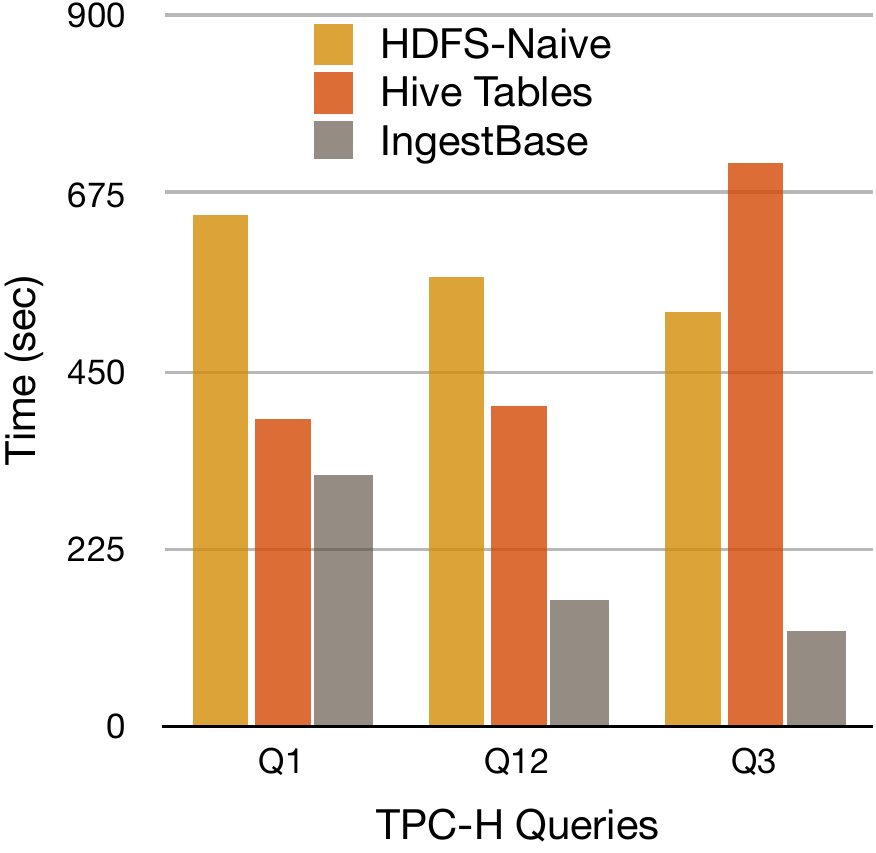}
\label{figure:tpch_queries}
}
\vspace{-0.4cm}
\caption{The effectiveness of ingestion-aware access methods in leveraging the ingestion logic at query time.}
\vspace{-0.5cm}
\end{figure*}

\subsection{Comparison with Hive Cooking Jobs}
\label{subsubsec:hive_baseline}

The typical practice is to prepare a dataset once it is already ingested into HDFS using query processing tools or MapReduce over HDFS.
Let us now compare \system with such an approach.

\noindent{\bf Setup.} We preload the data into HDFS and create an external Hive table to contain the data. We then run HiveQL queries to do three data transformations, namely functional dependency, denial constraint checking, and random sampling (these are the only transformations that are easy to represent in HiveQL).

\vspace{-0.2cm}
\begin{table}[h!]
\centering
\scriptsize
\begin{tabular}{| l | r | r | r |}
\hline
\textbf{Transformation} & \textbf{Hive} (s) & \textbf{\system} (s) & Improvement \\\hline\hline
Functional Dependency & 8,100 & 4,516 & 1.8x \\\hline
Denial Constraint & 2,616 & 1,011 & 2.6x \\\hline
Random Sampling & 2,274 & 371 & 6.1x \\\hline
\end{tabular}
\vspace{0.1cm}
\caption{Ingestion Overhead in Hive and \system.}
\label{table:overhead}
\vspace{-0.4cm}
\end{table}

\noindent{\bf Discussion} Table~\ref{table:overhead} shows the ingestion overhead, above the standard HDFS upload time, in Hive and \system. We can see that \system has $1.8\times$ less overhead than Hive for checking functional dependency and almost $6.1\times$ less overhead for random sampling. The reason is that \system piggy-backs these operations onto the ingestion process. For example, to generate random samples, \system incurs a single data read from disk while ingesting it into HDFS. Hive on the other hand, needs to re-read the entire dataset twice. Furthermore, it is tedious to run more complex transformations such as stratified sampling in Hive, and physical transformations such as erasure coding are not possible at all.
Thus, \system has utility both in terms of performance as well as flexibility in ingesting datasets in an ad-hoc manner.

\subsection{Ingestion-aware Data Access}

We now look at the performance of ingest-aware access plans using \system. We ran these experiments on the same cluster with the TPC-H dataset at scale factor $100$.
We consider the four typical query operators: (1)~projection, (2)~selection, (3)~aggregation, and (4)~join. 
For these experiments, we used the data access methods of \system in Hadoop MapReduce. 
We saw similar gains with Spark as well.

\noindent\textbf{Projection.} Figure~\ref{figure:operator_projections} shows the performance of \system{} and standard HDFS (HDFS-Naive) access methods on projection queries.
For \system, we consider projection over six different formats: string\footnote{The string serializer treats each line in the input as an unparsed record, i.e.,~the line is not parsed even at query time.} (\system~(str)), binary (\system~(bin)), column-grouped (\system~(cg)), PAX (\system~(pax)), RCFile (\system~(rcf)), and compressed PAX (\system~(cpax)).
HDFS-Naive parses the input tuple and emits the projected attributes in the map function. 
We varied the number of projected attributes from $1$ to $15$. 
As expected, Figure~\ref{figure:operator_projections} shows that \system{} allows to create optimized data layouts which could then be accessed efficiently from the MapReduce processing engine.

\noindent\textbf{Selection.} Figure~\ref{figure:operator_selections} shows the runtimes of \system and HDFS-Naive access methods on a selection query.
\system{} access methods provides several ways to filter the data, including post-filtering the data after reading it~(\system~(PoF)), late materializing only the qualifying rows of projected columns~(\system~(LM)), index accesses within each data block~(\system~(IA)), filtering data files~(\system~(PaF)), and filtering a fully sorted dataset~(\system~(GS)). HDFS-Naive applies the selection predicate in the map function.
We vary the query selectivity from $0.1$ ($10\%$) to $1.0$ ($100\%$). 
Again, we can see that \system{} access methods allow to pre-filter the relevant data effectively, thereby leading to faster query performance.

\noindent\textbf{Aggregation \& Join.} Figure~\ref{figure:operator_aggregation} shows an aggregation query over HDFS-Naive and \system access methods with key cardinalities of $1M$ and $150M$ tuples. 
For \system, we co-group the data and perform hash-based (\system~(Hash)) or sort-based (\system~(Sort)) aggregation. 
HDFS-Naive shuffles the data on the group-by key and computes the aggregate in the reducer.
Co-grouped data access naturally leads to better performance.
Figure~\ref{figure:operator_joins} shows the join query runtimes for HDFS-Naive and \system. \system co-groups the relations on the join keys and performs hash-join. HDFS-Naive shuffles the data on the join keys and produces the join result in the reducer. 
Again, the co-grouped data access results in better performance, as expected.

\noindent\textbf{TPC-H Queries.} Finally, Figure~\ref{figure:tpch_queries} shows the performance of HDFS-Naive, Hive tables, and \system access methods on three unmodified TPC-H queries: Q1, Q12, and Q3, involving $1$, $2$, and $3$ TPC-H tables respectively. We can see that applying custom ingestion logic and then performing ingestion-aware data access leads to better overall runtimes, compared to both to HDFS-Naive as well as Hive tables.

In summary, these experiments show that application-specific ingestion plans combined with ingestion-aware data access can be used to easily build 
high performance, novel query processing tools
over HDFS datasets.

\subsection{Fault Tolerance}

Recall \system allows users to express how to recover their datasets in case of failures.
In this section, we evaluate the efficiency of \system at recovering from failures. 

\noindent\textbf{Setup.} We killed one of the data nodes in HDFS and measured the time that \system took to recover the missing data files on that node.
We evaluated two recovery implementations in \system: (i)~\textit{replication based}, i.e.~simply increasing the replication factor of an equivalent data replica, and (ii)~\textit{transformation based},~i.e.~copy a data replica and transform it into the data representation of the missing data file.

\noindent\textbf{Discussion.} Table~\ref{table:ftoverhead} shows the per-file ($64$ MB) recovery overhead of \system.
\system takes a few milliseconds to recover a missing data file in replication-based recovery. This is because HDFS takes care of the entire recovery process. For transformation-based recovery case, \system takes  $\sim$4 seconds to recover a missing data file, even though it has to transform the data representation.

Thus, we see that \system preserves the fault-tolerance properties and quickly recovers missing or corrupted data files in case of failures. 
This high efficiency results from the fact that \system fully leverages HDFS fault-tolerance and simply points to the equivalent data files (with transformation if needed).

\begin{table}[h!]
\centering
\scriptsize
\begin{tabular}{| l | r | r | r |}
\hline
\textbf{Recovery Implementation} & Overhead (ms) \\\hline\hline
Replication-based & 18 \\\hline
Transformation-based & 4,140 \\\hline
\end{tabular}
\vspace{0.2cm}
\caption{Per-file Recovery Overhead in \system.}
\label{table:ftoverhead}
\vspace{-0.2cm}
\end{table}

\section{Related Work}
\label{section:relatedwork}

The traditional practice in database systems is to use so-called ``extract-transform-load'' (ETL) tools for transforming datasets from one database to another. However, ETL tools typically deal with high-level schema integration of data from different sources, or schema to schema transformations in SQL-like languages. In contrast, \system deals with low-level physical transformations of bytes. Several works have proposed high level languages for describing the physical data layouts. For example, GMAP proposed a data definition language to define the storage structures~\cite{gmap}; RodentStore presented a storage algebra to specify the way in which data should be laid out on disk~\cite{rodentstore}; OctopusDB~\cite{octopusdb} proposed storage views to create different physical representation of a logical log; WWHow!~\cite{wwhow} proposes to consider what, where and how as the three dimensions of data; and other researchers have proposed a language to easily specify rich storage constraints~\cite{constraineddesign}. However, these systems have focused on high level specifications in relational systems, rather than an end-to-end ingestion system. Still, \system could be used as a mechanism to apply these declarative languages.


A number of projects have focused on collecting, aggregating, and moving data into a single store engine, such as Flume~\cite{flume}, Sqoop~\cite{sqoop}, Gobblin~\cite{gobblin}, and Skool~\cite{skool}. 
However, all these systems were designed to move data from a single or several data sources to a specific target data store.
The Google Cloud Dataflow system aims at providing a unified programming model for developing and executing a wide range of data processing tasks, including a data upload task~\cite{dataflow}.
However, this system does not provide a declarative language and hence users must write code to build their data upload pipelines.
Furthermore, to the best of our knowledge, it does not provide fault-tolerance either.
While \system allows users to achieve the same, it additionally abstracts the data ingestion process and allows users to specify their data ingestion tasks declaratively.

Finally, as noted in the introduction, many researchers have proposed optimized data storage layouts in the context of MapReduce~\cite{hadooppp,starfish,rcfile,llama,trojanlayout}. However, these efforts have generally been designed to operate at a very low-level, involving modifications to HDFS to fit their needs. \system, on the other hand, is a more general data ingestion engine designed to ingest user datasets in an ad-hoc manner without requiring deep changes to the storage or compute substrate.
\section{Conclusion}
\label{section:conclusion}




Big data applications have fast arriving data that needs to be ingested quickly and for application-specific needs.
While a plethora of query processing tools have been proposed to arbitrarily query the data, there is a lack of ingest processing tools to arbitrarily prepare the data.
In this paper, we proposed \system{}, a declarative data ingestion framework that gives developers full control over how their datasets are preprocessed and ingested into a storage system.
We introduced the notion of \textit{ingestion plans}, which specify a sequence of logical operations (ingestion operators) to be performed on raw data as it gets ingested. 
\system{} hides the ingestion complexity from the developers by exposing a declarative interface, allowing them to easily build sophisticated ingestion plans.
These plans are later tuned using a rule-based optimizer and efficiently executed using a fault-tolerant runtime engine.
We showed through extensive experiments that \system provides good ingestion performance over several use-cases. 
In addition, ingestion-aware data access successfully leverages the preprocessing done during ingestion.
Although we experimented \system{} on top of HDFS, we believe the concept of ingestion plans can readily be extended to other distributed storage systems.


\vspace{-0.2cm}

%
%
%
%
%
%
%
%
%


%
%



%

\small
\bibliographystyle{IEEEtran}
\bibliography{IEEEabrv,references}

%
%

\end{document}